\documentclass[twocolumn]{aastex631}
\usepackage{graphicx} 

\begin{document}

\title{Characterizing the 3D Structure of Molecular Cloud Envelopes in the ``Cloud Factory" Simulations}

\author[0000-0003-0814-7923]{Elijah Mullens}
\affiliation{Space Telescope Science Institute, 3700 San Martin Drive, Baltimore, MD 21218, USA}
\affiliation{Department of Astronomy and Carl Sagan Institute, Cornell University, 122 Sciences Drive, Ithaca, NY 14853, USA}

\author[0000-0002-2250-730X]{Catherine Zucker}
\affiliation{Center for Astrophysics $\mid$ Harvard \& Smithsonian, 60 Garden St., Cambridge, MA, USA 02138}
\affiliation{Space Telescope Science Institute, 3700 San Martin Drive, Baltimore, MD 21218, USA}

\author[0000-0002-7743-8129]{Claire E. Murray}
\affiliation{Space Telescope Science Institute, 3700 San Martin Drive, Baltimore, MD 21218, USA}
\affil{Department of Physics \& Astronomy, Johns Hopkins University, 3400 N. Charles Street, Baltimore, MD 21218}

\author[0000-0002-0820-1814]{Rowan Smith}
\affiliation{School of Physics and Astronomy,
University of St. Andrews, North Haugh, St. Andrews, Fife KY16 9SS, UK}
\affiliation{Jodrell Bank Centre for Astrophysics, Department of Physics and Astronomy, University of Manchester, Oxford Road, Manchester M13 9PL, UK}

\begin{abstract}
We leverage recent numerical simulations of highly resolved star-forming regions in a Milky Way-like Galaxy to explore the nature of extended gaseous envelopes around molecular clouds. We extract a sample of two dozen star-forming clouds from the feedback-dominated suite of the ``Cloud Factory" simulations. With the goal of exploring the 3D thermal and chemical structure of the gas, we measure and fit the clouds’ radial profiles in multiple tracers, including $n_{\rm H_I}$, $ n_{\rm H_2}$, $n_{\rm H_{tot}}$, $n_{\rm CO}$, and gas temperature. We find that while solar neighborhood clouds recently detected via 3D dust mapping have radially symmetric, low-density envelopes that extend $\sim 10-15$ pc, the simulated cloud envelopes are primarily radially asymmetric with low-density envelopes that extend only $\sim 2-3$ pc. One potential explanation for the absence of extended envelopes in the simulated clouds may be the lack of magnetic fields, while a stronger local feedback prescription compared to solar neighborhood conditions may drive the radially asymmetric cloud morphologies. We make the pipeline to extract and characterize the radial profile of clouds publicly available, which can be used in complementary and future simulations to shed additional light on the key physics shaping the formation and evolution of star-forming structures in the Milky Way. 

\end{abstract}


\section{Introduction} \label{sec:intro}

An appreciable fraction of the mass budget of the interstellar medium (ISM) belongs to dense, cold formations termed molecular clouds \citep[e.g.,][]{Blitz_1993}. These molecular clouds are the sites of star formation in galaxies, and as a result play a key role in galaxy evolution. The structure of a molecular cloud is subject to a complicated interplay between galactic dynamics, self-gravity, feedback, and magnetic fields \citep[e.g.,][]{Dobbs_2014}. Understanding the structure of molecular clouds is therefore integral not only to understanding these physical mechanisms, but also understanding where, when, and how stars form and interact with their broader galactic environment \citep[e.g.,][]{life_and_times}.

Molecular cloud structure has been extensively studied using both observations and simulations. Traditionally, observations of molecular cloud structure have largely been limited to either 2D projected space \citep[via integrated plane-of-the-sky dust emission or dust extinction maps;][]{Lombardi_2010, Planck_2011, Andre_2010, Froebrich_2010}, or 3D {\it ``position-position-velocity"} space obtained via carbon monoxide (CO) spectral-line mapping \citep{Ridge_2006, Dame_2001, Jackson_2006, Duarte_Cabral_2021, Colombo_2019}, where the third axis is the radial velocity of the gas obtained from the Doppler effect, not distance.   

However, with the launch of \emph{Gaia}, true 3D spatial, {\it ``position-position-position"}  maps of the interstellar medium can now be reconstructed with the distance resolution necessary to resolve the internal structure of molecular clouds, owing to a technique known as 3D dust mapping \citep{Green_2019, Vergely_2022, Dharma_2023, Rezaei_2023}. Notably, \citet{dust_maps} produce a highly resolved $\sim$2 pc resolution 3D dust map of the solar neighborhood (out to a distance of $\sim$ 400 pc from the Sun) using distance and extinction estimates to a large number of stars inferred from a combination of \emph{Gaia} DR2 astrometric data and \emph{Gaia}, 2MASS, Pan-STARRS, and ALLWISE photometric data \citep[see][]{Anders_2019}. 

\citet{Zucker2021} utilize the 3D dust map from \citet{dust_maps} to create 3D models of local molecular clouds and characterize their 3D spatial structure.\footnote{See also \citet{Dharma_2023} for another work examining the 3D spatial structure of local molecular clouds, and Figure 6 in \citet{Cahlon_2024} for a comparison between the \citet{Dharma_2023} and \citet{dust_maps} results.} \citet{Zucker2021} find that local molecular clouds are filamentary, and extract radial volume density profiles from the 3D dust map to study the extended structure of clouds as a function of distance from their filamentary ``spines". \citet{Zucker2021} find that the averaged radial volume density profiles of local molecular clouds are best fit with a two-component Gaussian function, where an inner Gaussian fits the higher-density peak near the core of the cloud and an outer Gaussian fits the lower-density extended tail. \citet{Zucker2021} propose that the two separate Gaussians either represent a chemical phase transition (between low-density atomic gas in the outskirts and dense molecular gas near the core of the cloud) or an atomic gas thermal phase transition (between the Unstable Neutral Medium near the outskirts of the cloud and the Cold Neutral Medium near the core of the cloud). However, observational 3D dust data only probes differential extinction which can only approximate the total hydrogen gas volume density within the cloud thanks to constraints on the  wavelength-dependent extinction curve \citep{Draine_2009}. Any information on gas phase, composition, and temperature is not directly accessible from 3D dust maps alone. In order to test the hypothesis from \citet{Zucker2021}, we instead turn to simulations of molecular clouds.

Simulations of molecular clouds over the last decade have progressed from isolated clouds with only one or two feedback mechanisms to simulations that follow molecular cloud lifetimes in full with both local and galactic contexts included. One goal of a specific subset of simulations is to elucidate the relationship between molecular cloud structure and its chemical composition as the cloud evolves over time and forms both dense gas and young stars. These simulations have tied together chemical processes that form molecular hydrogen ($\rm H_2$) and carbon monoxide (CO) with the large-scale hydrodynamic evolution of molecular clouds \citep[e.g.][]{sim_AREPO, Sim_GIZMO,sim_Post_Processing,Sim_SILCC-Zoom,Sim_New_Tigress,Sim_GIZMO}.

In particular, the Cloud Factory simulations of \citet{cloud_factory} study the effects of the galactic potential, gravitational collapse, and supernova feedback on the formation and destruction of $\rm H_2$ and CO. More broadly, 3D simulations like the Cloud Factory track a network of filamentary clouds and their chemical compositions, which provide a large set of simulated clouds to compare with multi-wavelength observations probing the chemical and thermal structure of Milky Way clouds \citep[see][for a more complete review of resolved cloud simulations]{life_and_times}.

With the advent of \emph{Gaia} and 3D dust mapping, there is now sufficient observational data on the 3D structure of molecular clouds to facilitate a comparison to 3D simulated filamentary clouds. In this paper, we extend the methodology of \citet{Zucker2021} and develop a pipeline that efficiently compares the 3D structure detected in resolved numerical simulations of star-forming clouds to new 3D observational data. Leveraging this pipeline we aim to extract the radial profiles and morphologies of star-forming clouds in the ``Cloud Factory" simulations to quantitatively examine differences in cloud structure detected in observations and simulations. If the simulations quantitatively reproduce the observations, the pipeline is built to examine whether the two-component Gaussian structure of molecular clouds identified in previous 3D dust mapping observational work may be tracing a chemical or thermal phase transition.

In \S \ref{sec:cloudfactory}, we summarize the Cloud Factory suite of simulations utilized in this work. In \S \ref{sec:Methods}, we detail the methods used to analyze the simulations. In \S \ref{sec:results}, we summarize the results of the pipeline applied to the Cloud Factory simulations and highlight the range of structures detected in an ensemble of the clouds. In \S \ref{sec:discussion} we discuss the results and propose potential explanations for discrepancies between simulations and observations.  Finally, we conclude in \S \ref{sec:conclusion}. 

\section{The Cloud Factory} \label{sec:cloudfactory}

The aim of the Cloud Factory simulations is to investigate the formation of cold dense structure in the interstellar medium while including the galactic scale forces thought to be responsible for cloud assembly. Specifically the simulations contain an analytic galactic potential, and therefore naturally have disks with differential rotation and well-defined spiral arms. Full details can be found in \citet{cloud_factory} but we summarize here.

The models are performed using the \textsc{Arepo} code \citep{Springel10} with custom physics modules to describe star formation and cold, dense gas. The chemical evolution of the gas is modelled using the hydrogen chemistry of \citet{Glover07a,Glover07b}, together with the highly simplified treatment of CO formation and destruction introduced in \citet{Nelson97}. Our modeling of the hydrogen chemistry includes H$_2$ formation on grains, H$_{2}$ destruction by photo-dissociation, collisional dissociation of atomic hydrogen, H$^{+}$ recombination in the gas phase and on grain surfaces (see Table 1 of \citealt{Glover07a}), and cosmic ray ionization. We assume that the strength and spectral shape of the ultraviolet portion of the interstellar radiation field (ISRF) are the same as the values for the Solar neighborhood derived by \citet{Draine78} (equivalent to $1.7\times$ the field strength derived by \citealt{Habing68}). To treat the attenuation of the ISRF due to H$_{2}$ self-shielding, CO self-shielding, the shielding of CO by H$_{2}$, and by dust absorption, we use the \textsc{TreeCol} algorithm developed by \citet{Clark12b} assuming a shielding length of $L_{\rm sh} = 30 \: {\rm pc}$, We adopt a cosmic ray ionization rate of $\xi_{\rm H} = 3 \times 10^{-17}$ s$^{-1}$ for atomic hydrogen, and a rate twice this for molecular hydrogen. Finally we assume a Solar metal abundance, and a 100:1 gas-to-dust ratio. Heating and cooling of the gas is computed simultaneously with the solution of the chemical rate equations.

Star formation is modeled via sink particles \citep{Bate95}, which are non-gaseous particles that represent collapsing regions of gas that will form small (sub)clusters of stars. These are formed by checking if regions of gas exceed a critical density and are unambiguously bound, collapsing, and accelerating inwards. \textit{Only} if these criteria are met will the gas be replaced with a sink particle, which can then accrete additional mass that falls within a chosen accretion radius of the cell if it is gravitationally bound to it. 

Using the model of \citet{Sormani17}, we sample the IMF and associate supernova with the massive stars as described by \citet{Tress20a}. For each supernova, we calculate an injection radius, which is the radius of the smallest sphere centred on the supernova that contains at least 40 grid cells. If the injection radius is smaller than the expected radius of a supernova remnant at the end of its Sedov-Taylor phase, we inject thermal energy from the supernova; otherwise, we inject momentum \citep[e.g.][]{Gatto15}. Mass is returned with each supernova explosion such that when the last supernova occurs the gaseous component of the sink is exhausted. The sink is then turned into a star particle. To account for type Ia supernova, we also randomly select a star particle every 250 years and create a supernova event at its position.

The gravitational potential of non-gaseous cells is determined using an analytic potential. For the axisymmetric part of the potential, we use the best fitting model of \cite{McMillan17}, which was created to be consistent with various observational and theoretical constraints for the Milky Way and consists of the sum of a bulge, disk, and halo component. We then include a spiral perturbation to the potential, generated in the same way as in \citet{Smith14a}. Briefly, we use a four-armed spiral component from \citet{Cox02} with a pitch angle $\alpha = 15^\circ$ and a pattern speed of $2 \times 10^{-8} \, {\rm rad \, yr^{-1}}$.

The initial condition is inspired by the Milky Way gas disk model of \citet{McMillan17}, which is based on a combination of observational constraints and theoretical modelling. It consists of two density distributions for the HI and H$_2$ that decline exponentially at large radii. As we focus on Milky Way-like clouds outside the central bar, we neglect galactic radii smaller than 4 kpc. The gas disk is given the initial rotation curve that arises from the analytic potential, which for our disk corresponds to a rotation curve of order 220 kms$^{-1}$.

The simulations initially have a base mass resolution of 1000 M$_{\odot}$. However once a steady state has been reached (after 150 Myr) we turn on refinement for two spiral arm passages ($\sim 70$ Myr) within a 3 kpc box that co-rotates with the gas centred on a galactic radius of 8 kpc. In this high resolution region the gas has a target mass of initially 100 M$_{\odot}$ for the first 60 Myr, but it is further lowered to 10 M$_{\odot}$ \ for the final 10 Myr. In addition to the mass requirement we always require that the Jeans length is resolved by at least four cells up to our sink creation density. Within the 3 kpc box, sinks form above a minimum creation density of 574 cm$^{-3}$ (typically about $10^4$ cm$^{-3}$ in practice once energy checks are satisfied) and have an accretion radius of 0.1 pc. To avoid discontinuous jumps in the cell size, particularly where the target resolution is changing at the boundaries of the high resolution box, we require that the cell radius of adjacent \textsc{Arepo} cells can differ by no more than a factor of two at any time throughout the entire simulation volume.

The Cloud Factory simulations proceed to refine to even smaller scales in selected clouds, but for this study we are interested in the statistics of a large area and so we use only the 3 kpc box. ‘The original models also had two versions of this 3 kpc box --- a ``potential-dominated version" and a ``feedback-dominated version". The potential-dominated version is where gas self-gravity was only turned on for the life-times of current molecular clouds (5 Myr), and supernova feedback is random. The feedback-dominated version is where gas self-gravity has been turned on for multiple molecular cloud generations (50 Myr) and supernova feedback is tied to the sink particles that represent star formation. The aim of the two versions is to focus on the differences arising between two extreme cases. In this paper we focus primarily on the `feedback-dominated' case, as it more faithfully reproduces the larger scale-height of the disk in the solar neighborhood seen in recent 3D dust maps \citep{dust_maps, Gordian2023}. We emphasize, however, that the feedback-dominated case is likely an upper limit on the strength of supernova feedback experienced in the ISM. We additionally take a 0.5 kpc subset of the z direction since the z direction in the 3 kpc box is largely empty. We revisit the `potential-dominated’ case in \S \ref{sec:discussion}.’

\section{Methods}\label{sec:Methods}

Here we outline our methodology for characterizing the structure of molecular clouds in the Cloud Factory simulations.  We start by sub-dividing the Cloud Factory simulations into smaller sub-grids and identifying sub-grids with an appreciable fraction of molecular star-forming gas (\S \ref{subsec:converting-sorting-grids}). Then, building on the methodology of \citet{Zucker2021}, we iteratively extract and ``skeletonize" simulated molecular clouds in the simulations (\S \ref{sec:skeletonization}), compute their radial profiles (\S \ref{sec:pipeline}), and fit the radial volume density profiles with one- and two-component Gaussian functions (\S \ref{subsec:fitting_data}). After extracting the clouds and characterizing their radial profiles, we also sort clouds based on whether they contain sink particles, in order to correlate their observed cloud structure with their degree of star formation. 

\subsection{Converting and Sorting Grids}\label{subsec:converting-sorting-grids}
As discussed in \S \ref{sec:cloudfactory}, we focus primarily on the feedback-dominated suite of the Cloud Factory simulations. However, we repeat the analysis detailed in the following sections on one grid of the potential-dominated suite and report preliminary results in \S \ref{sec:discussion}. We start by dividing the feedback-dominated box (a $3 \times 3 \times 0.5$ kpc slab) into sub-regions. Each sub-region is a $500 \times 500 \times 500 \; \rm pc$ box and has been regridded to a uniform 1 $\rm pc^3$ resolution. The Cloud Factory is built on the AREPO code \citep{Arepo_1,Arepo_2}. The Cloud Factory implementation of AREPO tracks the total gas density ($\rho_{\rm gas}$) and temperature, as well as the carbon monoxide abundance ($x_{\rm CO}$), molecular hydrogen abundance ($x_{\rm H_2}$), and ionized hydrogen abundance ($x_{\rm H^+}$) with respect to the total gas density. 

We convert AREPO units to physical units and derive the total volume density of hydrogen nuclei using the mean molecular weight of hydrogen ($n_{\rm H_{\rm tot}} = \frac{\rho_{\rm gas}}{1.4 \times m_p}$, where 1.4 accounts for the helium abundance). Using the total gas density and abundance grids we then derive the volume density of carbon monoxide, molecular hydrogen and ionized hydrogen ($n_{\rm CO,H_2,H^+} = x_{\rm CO,H_2,H^+} \times n_{\rm H_{tot}}$). Atomic hydrogen ($\rm H_I$) volume density is derived by subtracting the contributions of the molecular and ionized hydrogen from the total hydrogen gas density ($n_{\rm H_I} = n_{\rm H_{tot}} - 2n_{\rm H_2} - n_{\rm H^+}$). We then narrow down the grids based on the amount of molecular gas contained within them. We choose to only analyze grids containing at least 1000 voxels ($\rm > 1000 \; pc^3$ volume, equivalent to a spherical cloud with a radius of $\approx 6$ pc) with a relative abundance $x_{\rm H_2} > 0.4$. Grids with appreciable fractions of molecular gas provide the best comparison to observed molecular clouds, and are utilized in the remainder of the analysis. We identify nine grids that fit this criterion, namely x1465y1390z1400, x1465y1400z1395, x1465y1400z1400, x1465y1405z1395, x1465y1400z1400, x1470y1400z1395, x1470y1400z1400, x1475y1390z1395, and
x1475y1390z1400.

\subsection{Simulated Cloud Identification and Skeletonization}\label{sec:skeletonization}

\begin{figure*}[ht!]
\includegraphics[width=1.0\textwidth]{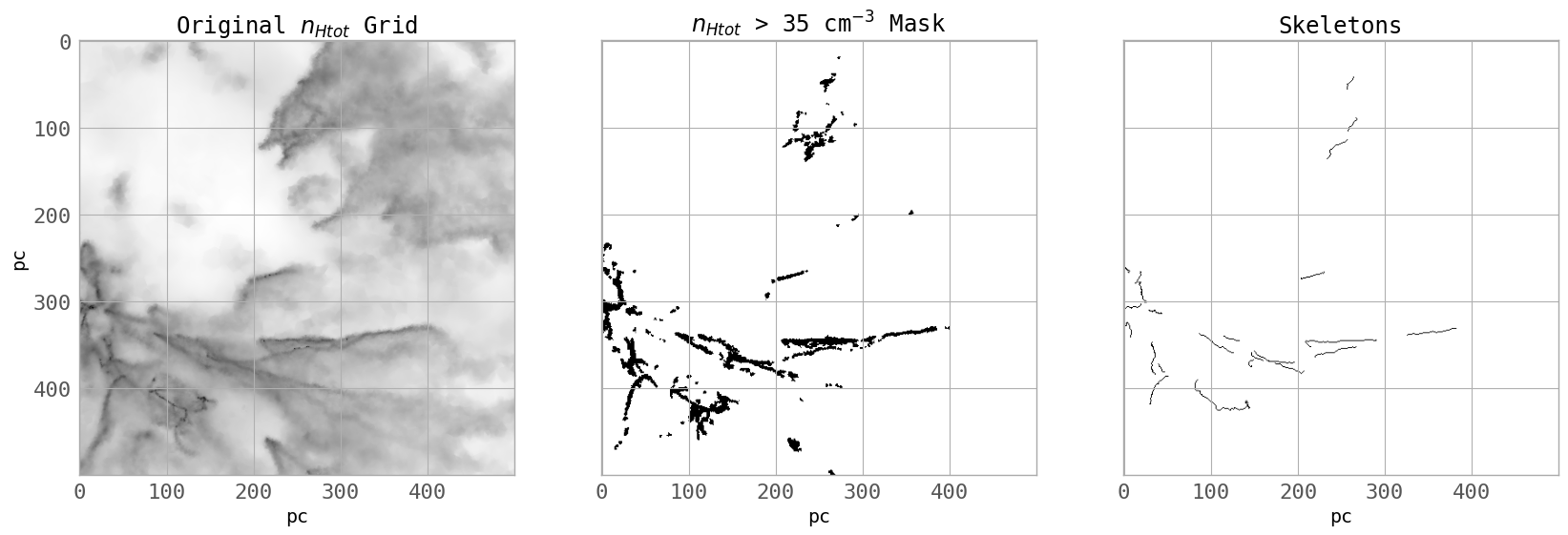}
\caption{Example of cloud identification and skeletonization for a single simulation sub-grid. The 3D grid is displayed with one axis collapsed. The first panel displays the total volume density of hydrogen nuclei in the grid. The second panel displays the same simulation grid after applying a density threshold mask of $n_{\rm H_{tot}} > 35 \; \rm cm^{-3}$, which is used to identify potential clouds. The third panel displays the resulting topological skeletons. Small clouds present in the second panel are removed from the dataset and are not skeletonized. (grid x1465y1400z1400) \label{fig:glue_example}}
\end{figure*}

In previous work molecular clouds and their sub-structure have been characterized by topological skeletons \citep{filfinder, getfilaments}. Topological skeletons are one-pixel-wide representations of two- or three-dimensional shapes that are equidistant to the boundary surface of the original shape. Topological skeletons are derived by starting from the boundary of a shape and subtracting off one pixel, moving inwards until only a single pixel-wide shape remains. For example, the skeletonization of a two dimensional circle will result in a single point in the center, and the skeletonization of a three dimensional cylinder will result in a straight line through the center of the cylinder. 

Following the formalism of \citet{Zucker2021} we compute topological skeletons of simulated molecular clouds by employing a 3D version of the FilFinder package \citep{filfinder}. We first apply a density threshold mask of $n_{\rm H_{tot}}>35 \; \rm cm^{-3}$ to the entire simulation grid, following the same line of reasoning as \citet{Zucker2021}. Individual features that survive masking are then iteratively identified. This is accomplished by looping over each voxel in the grid that survive thresholding and then flooding the regions surrounding those voxels. If a flooded feature contains more than 1000 non-empty voxels, then morphological closing via dilation and erosion is performed to fill small holes. The filling of small holes ensures the skeletonization can be performed efficiently. If a feature has less than 1000 voxels, that feature is removed from future cloud-seeking iterations. This ensures that small, spurious features are not analyzed. Significant features are then skeletonized using FilFinder. Skeleton length and total feature mass are derived, and if a skeleton's total length is greater than 20 pc the skeleton is kept for future analysis. A 20 pc minimum length is chosen as the approximate lower bound due to the range of lengths observed for molecular clouds in the solar neighborhood \citep{Zucker2021}. See Figure \ref{fig:glue_example} for a visual of the cloud identification and skeletonization process.

The simulated clouds can also be sorted based on whether or not they are forming stars. We flag star-forming clouds by iterating over each cloud and determining if there are any ``star-forming sinks" [see sink particles introduced in \S \ref{sec:cloudfactory}] inside the cloud mask. If a cloud contains any star-forming sinks it is labeled as a star-forming cloud. This specification allows for a better cross-comparison between subsets of simulated clouds and the observed clouds in \citet{Zucker2021} which are all star-forming. Specifically, \citet{Zucker2021} analyzed a subset of clouds from the Star Formation Handbook \citep{Northern_Handbook, Southern_Handbook}, which summarizes the sixty most important star-forming regions within 2 kpc of the Sun. We refer readers to \citet{Northern_Handbook, Southern_Handbook} for more information on the star-forming properties of the observed clouds.

\subsection{Simulated Cloud Analysis Pipeline}\label{sec:pipeline}

\begin{figure*}[h!]
     \centering
         \includegraphics[width=1.0\linewidth]{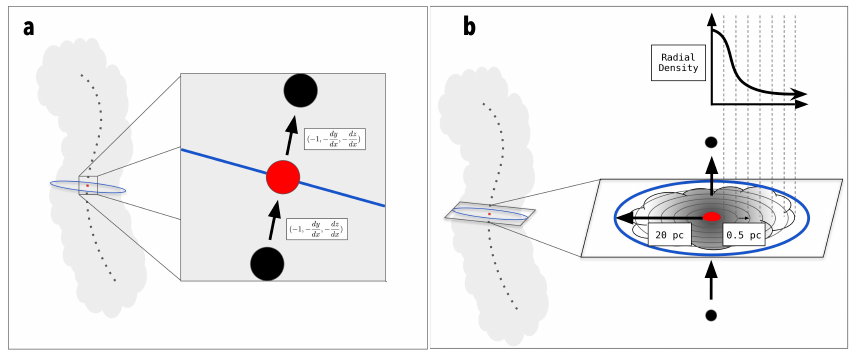}
     \caption{Graphical demonstraion of the analysis pipeline. (a) Zoom-in displaying the normal vector pointing between skeleton points. This normal vector is used to define a plane that bisects the cloud. (b) Zoom-in display of the 20 pc `slice' around a single skeleton point, which is then split into 0.5 pc radial bins and used to derive an average radial density profile for one skeleton point.}
     \label{fig:Point_by_point example}
\end{figure*}

\begin{figure*}[h!]
\centering
\includegraphics[width=0.7\textwidth]{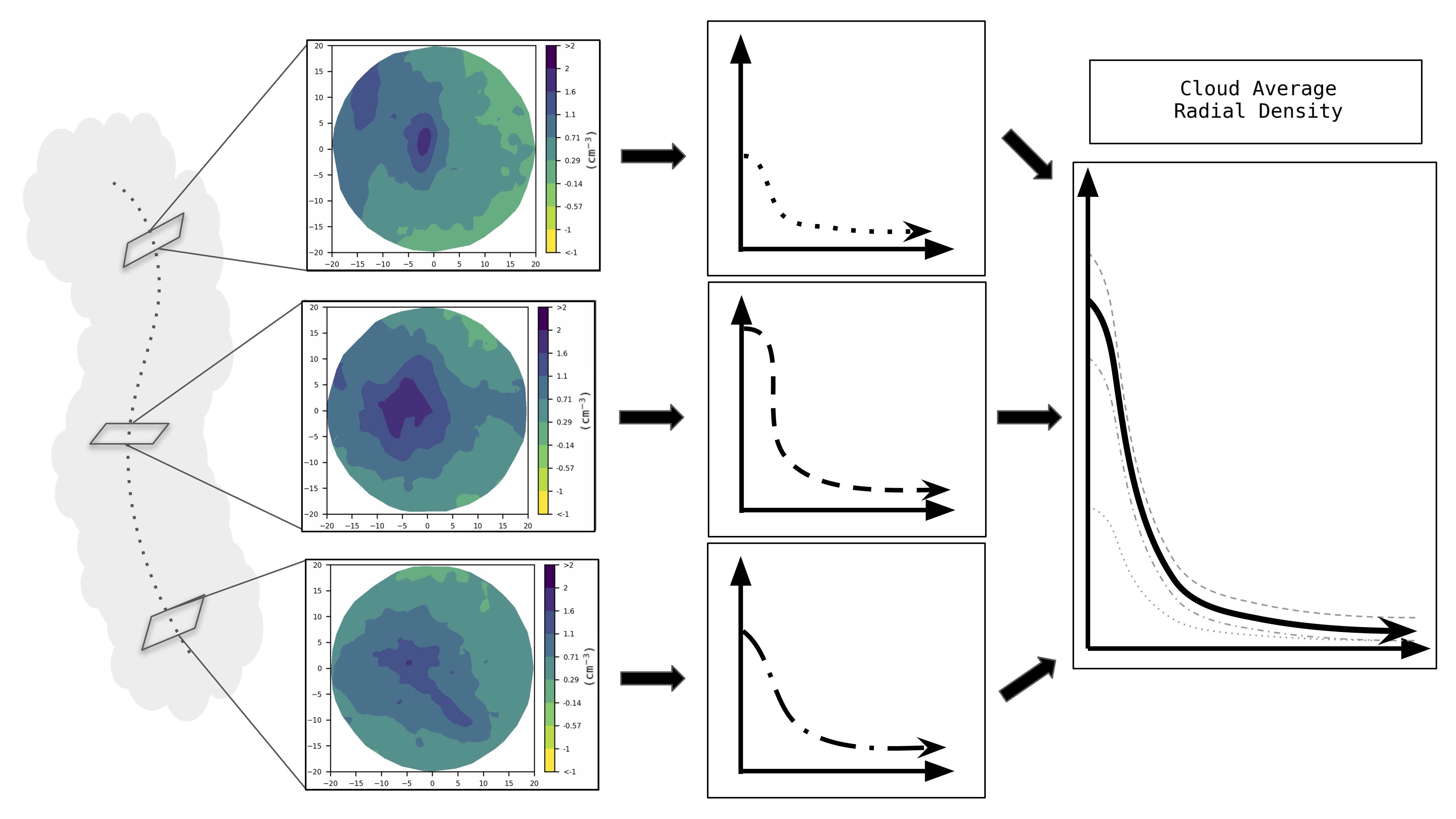}
\caption{Deriving the cloud average radial volume density profile for an entire cloud. Density slices are computed for each skeleton point. The radial volume density profile for each skeleton point is derived and then averaged to construct a cloud averaged radial volume density profile. Slices shown are from an observed molecular cloud \citep[Perseus;][]{Zucker2021} and individual radial volume density profiles are exaggerated.}
\label{fig:Cloud Average}
\end{figure*}


Once we extract a set of skeletons following the procedure outlined in \S \ref{sec:skeletonization}, we then calculate normal vectors between skeleton points in order to define cloud-bisecting planes to derive radial profiles. This is accomplished by fitting a linear spline to the skeleton points (a function that connects individual points with piece-wise linear functions), from which the first-order cartesian derivatives for each point in the skeleton are derived. These derivatives are used to define a normal vector pointing from the current point being analyzed to the next ordered point in the skeleton via $(-1, -\frac{dy}{dx},-\frac{dz}{dx})$. This normal vector is used to define a circular two-dimensional plane with a radius of 20 pc that cuts through the current point being analyzed and is orthogonal to the direction of the skeleton (see Figure \ref{fig:Point_by_point example}).

An integral difference to the analysis done in \citet{Zucker2021} and this work is that the simulation grids have access to more information than the dust density that was converted to the total hydrogen gas density using a wavelength-dependent extinction curve. The slices defined herein are purely geometric in cartesian space and do not encode any density or temperature information. We apply the slices to the simulation grid in order to extract and interpolate information on the density ($n_{\rm H_{tot}}$, $n_{\rm CO}$, $n_{\rm H_2}$, $n_{\rm H_I}$, $n_{\rm H^+}$) and gas temperature ($T_{\rm gas}$). We then derive radial profiles for each tracer. On a slice-by-slice basis the radial profile for each tracer is computed. We do this by first defining 40 radial distance bins (radius of $0-20$ pc extending from the skeleton point with steps of 0.5 pc). For each radial distance bin we compute the median of the density and temperature values (see Figure \ref{fig:Point_by_point example}). This process is repeated for each skeleton point and results in a set of radial profiles for each skeleton point in a cloud. Finally, each individual radial profile is averaged to obtain the cloud averaged radial volume density profile in each chemical tracer as well as temperature profiles (Figure \ref{fig:Cloud Average}). In this work, only the skeleton points which have core values (taken from the first radial bin) greater than the density threshold mask ($n_{\rm H_{tot}}>35 \; \rm cm^{-3}$) are included in the computation of the cloud average radial profiles to ensure we are not including spurious low-density patchy regions which may dilute the averaged profile.

The total gas density ($n_{\rm H_{tot}}$) radial profiles can be used directly as a basis of comparison with observations, since all 3D observational data from \citet{Zucker2021} is converted from the native units of the 3D dust map to $n_{\rm H_{tot}}$. If a simulated cloud's total gas density radial profile is comparable to observations, a mapping can be formulated from $n_{\rm H_{tot}}$ to the chemical and temperature profiles. This mapping can be then applied to observed radial profiles to test whether or not a chemical or temperature phase transition is occurring within a cloud. 

Slices that bisect the cloud serve a dual purpose. In addition to providing a tool by which radial profiles can be extracted, they also provide a visual on the morphology of a cloud. Comparative analysis between simulated cloud and observed cloud 3D morphology can be advantageous for diagnosing discrepancies between simulations and observations. The methodology above computes radial profiles by taking the median across radial bins, a process which relies on the assumption of radial symmetry. If a cloud is highly asymmetric, taking the median of density values in a radial bin will not accurately capture cloud structure. Additionally, if simulated clouds' average radial density profiles for the total gas density are not matching observations, there is no clear method by which to diagnose potential differences between simulated clouds and observations. To provide a tool to visually diagnose why simulated clouds may not be consistent with observations, the pipeline also outputs slices in each tracer and converts them into animations for visual analysis.

\subsection{Fitting Data}\label{subsec:fitting_data}

Once we extract the cloud averaged radial density profiles for the sample in all tracers ($n_{\rm H_{tot}}$, $n_{\rm CO}$, $n_{\rm H_2}$, $n_{\rm H_I}$, $n_{\rm H^+}$,and $T_{\rm gas}$)\footnote{The pipeline is also designed to run on the dust temperature data, $T_{\rm dust}$. However, we do not consider $T_{\rm dust}$ for further analysis, since it is represents the effective cooling from the dust}, we perform single and multi-component Gaussian fits to the density tracers. We also perform sigmoid, as well as spline fits, to the gas temperature profiles. 

We fit the gas temperature radial profile of each cloud to investigate the radial distance at which atomic gas is mostly cold neutral medium (CNM), mostly unstable neutral medium (UNM), and mostly warm neutral medium (WNM). From \citet{CNM_transitions}, atomic gas is in the CNM phase at $20-500$ K and the WNM phase at $3000-8000$ K. We therefore use the temperature profile fit to place markers at $T_{\rm gas}\sim$250 K and $T_{\rm gas}\sim$3000 K as the temperatures where atomic gas is ``mostly" CNM and ``mostly" WNM. We tested both sigmoid and spline fits to the temperature profiles. Due to cloud asymmetries, we found that the average temperature profile had large scatter and that sigmoid fits were incorrectly placing temperature markers. We opt to utilize the spline fit to more confidently determine which radial distance corresponds to $T_{\rm gas}=250$ K and $T_{\rm gas}=3000$ K in the temperature radial profile.

We fit the one and two-component Gaussians to all density tracers. For the single-component Gaussian fit, our model is defined as:
\begin{equation}
n_X(r) = a \: \exp(\frac{-r^2}{2\sigma^2})
\end{equation}
where $n_X$ is the volume density of tracer $X$, $r$ is the radial distance (pc), $a$ is the amplitude (peak density, $\rm cm^{-3}$) and $\sigma$ is the standard deviation (pc). The one-component fits were performed with a Levenberg-Marquardt least squares fitter. For the two-component Gaussian fit, our model is defined as: 
\begin{equation}
n_X(r) = a_1 \: \exp(\frac{-r^2}{2\sigma_1^2}) + a_2 \: \exp(\frac{-r^2}{2\sigma_2^2})
\end{equation}
where $a_1,\sigma_1$ and $a_2,\sigma_2$ are the amplitudes and standard deviations of the inner and outer Gaussians, respectably. The two-component fits were performed utilizing a sequential least squares optimization algorithm. In \citet{Zucker2021} the first 2 pc are excluded from the fit since optical stellar photometric and astrometric measurements (the key ingredients in 3D dust mapping) become either scarce and/or unreliable in the core of clouds due to high levels of dust extinction. For simulations we have access to core information without having to consider dust extinction. However, in order to ensure the best basis of comparison with observations we exclude the first 2 pc from our fits as well. The results stay consistent regardless if we include or exclude the inner 2 pc. We emphasize that the cloud average radial $n_{\rm H^+}$ profile is always a poor fit since ionized hydrogen has a radial profile with a non-zero center, which we do not take into account with our fits and is beyond the scope of this work.

We found that the choice to use maximum number density of molecular gas and atomic gas to differentiate between molecular and atomic clouds was incorrectly flagging clouds. The atomic gas number density ($n_{\rm H_I}$) in the cores of Cloud-Factory simulated clouds is on par with the molecular gas number density ($n_{\rm H_2}$) (see the core density values in Appendix Figure \ref{fig:forward_model}). We decide to utilize the star-forming subset of simulated clouds as a better basis of comparison with observed clouds because all molecular clouds analyzed in \citet{Zucker2021} are known to be star-forming.  

The pipeline we describe here is publicly available and open-source on GitHub. We include two versions of the pipeline: a Cloud Factory specific version that loops over the Cloud Factory simulation grids, and a version that takes in a single 3D grid (observational or simulated) and performs the above analysis on whatever tracer the grid uses. Example notebooks and data are included so that simulators and observers can run a similar analysis on their own data. 

\section{Results}\label{sec:results}

\subsection{Ensemble Results}

\begin{figure*}[ht!]
\centering
\includegraphics[width=0.8\textwidth]{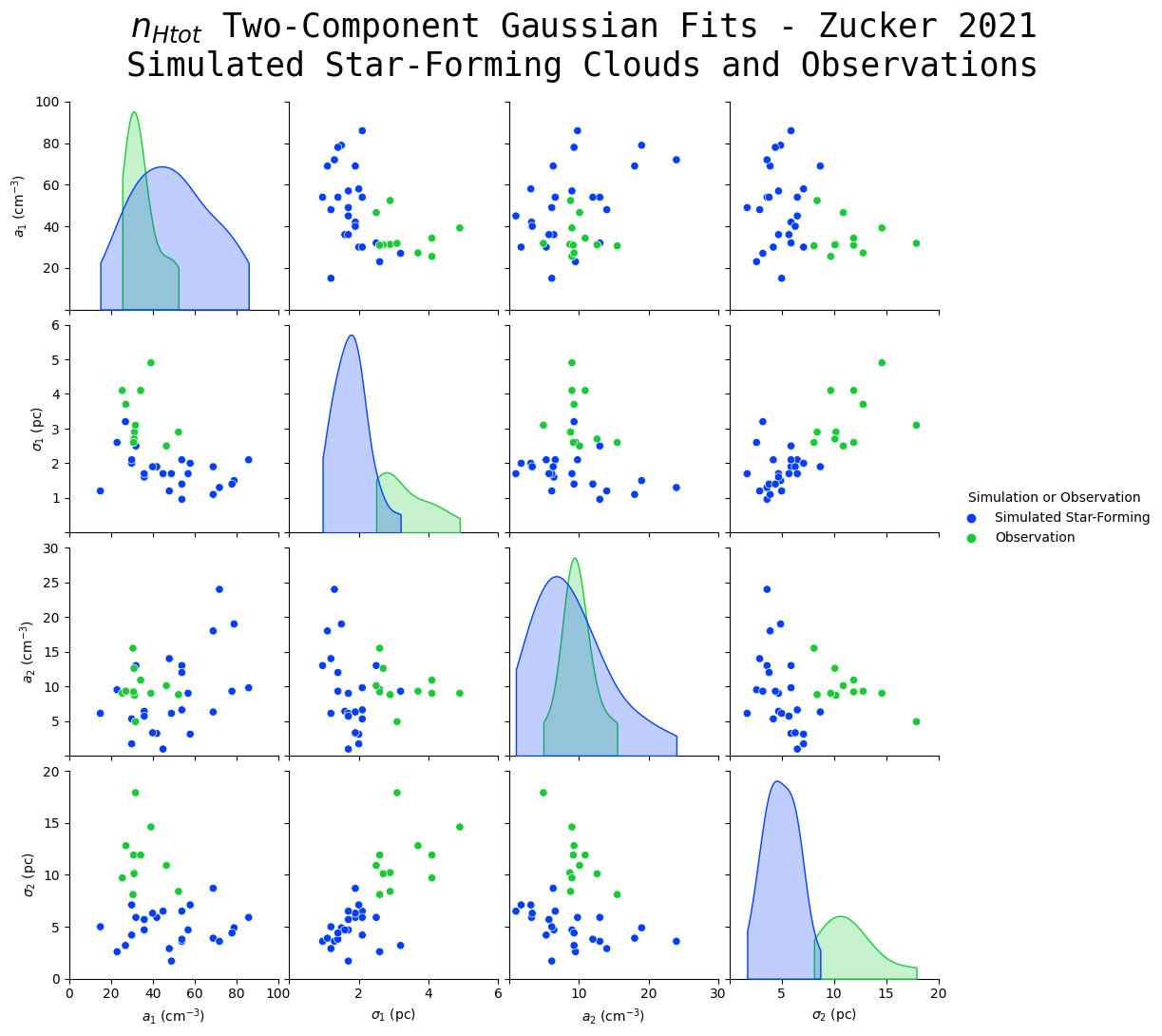}
\caption{Corner plot comparing the two-component Gaussian fits of observed clouds and simulated star-forming clouds. The diagonal shows a smoothed version of histograms comparing the two sets of clouds. Cloud peak densities ($a_1$, $a_2$) are largely consistent across the two data sets. However, cloud widths ($\sigma_1$, $\sigma_2$) show more significant disrepancies.}
\label{fig:corner_Zucker}
\end{figure*}

We applied the pipeline described in \S \ref{sec:Methods} to nine simulations sub-grids with an appreciable amount of molecular gas and analyzed 125 clouds in total with 36 being flagged as star-forming based on the presence of sink particles within their cloud masks. Out of the 36 star-forming clouds, only 24 had two-component Gaussian fits that successfully converged (see \S \ref{sec:individual-results} for details; eleven clouds failed to converge with $\sigma_{1,2}\sim 0$ pc while one cloud was an outlier with $a_1 \geq 100$ cm$^{-3}$). The ensemble results for the two-component Gaussian fits to the $n_{\rm H_{tot}}$ distributions of the 24 simulated star-forming clouds  are shown in Figure \ref{fig:corner_Zucker}, alongside the observed solar neighborhood clouds from \citet{Zucker2021}. The Gaussian peaks ($a_1$, $a_2$) for simulated clouds ($\sim$50 cm$^{-3}$, $\sim$8 cm$^{-3}$) agree mostly with observations ($\sim$30 cm$^{-3}$, $\sim$9 cm$^{-3}$) by construction: recall that the same density threshold that was used to define observational skeletons was used to find clouds in the simulations. Simulated clouds deviate from observations on the remainder of the two-component Gaussian fit parameters. The most striking difference is the inner and outer Gaussian widths ($\sigma_1,\sigma_2$) for simulated star-forming clouds and observations. The widths for both the inner and outer Gaussians of simulated star-forming cloud have lower values than observations. The Gaussian widths ($\sigma_1,\sigma_2$), extend out to $\sim$3 pc, $\sim$11 pc for observations but only extend out to $\sim$2 pc, $\sim$5 pc for simulations. 

Ensemble $n_{\rm H_{tot}}$ two-component and one-component Gaussian fits, as well as cloud length, mass, and $n_{\rm CO}$, $n_{\rm H_2}$, and $n_{\rm H_I}$ one-component Gaussian fits are highlighted in Table \ref{table:everything}. Modeling uncertainties in the simulated cloud radial profile fits is difficult; there are no intrinsic uncertainties stemming from the simulations and any `uncertainty’ in the reported fits would be due to the variations in the profiles on a slice-by-slice basis. We instead opt to report estimates of the population-level spread in the distribution of each reported value in Table \ref{table:population_statistics}.

The ensemble two-component gaussian fit results are the first indication that the simulated clouds are not aligning with observational expectations. In particular, the width of the first Gaussian $\sigma_1$ indicates that the inner peaks of the simulated cloud radial profiles are thinner than the observed radial profile peaks, and the width of the second Gaussian $\sigma_2$ indicates that the tails of the simulated cloud radial profiles are falling off much faster than the observed radial profile tails. In order to diagnose the discrepancy between simulation and observation, we analyze the radial profiles and morphologies of specific clouds in detail.

\subsection{Individual Cloud Examples}\label{sec:individual-results}

\begin{figure*}[ht!]
\includegraphics[width=\linewidth]{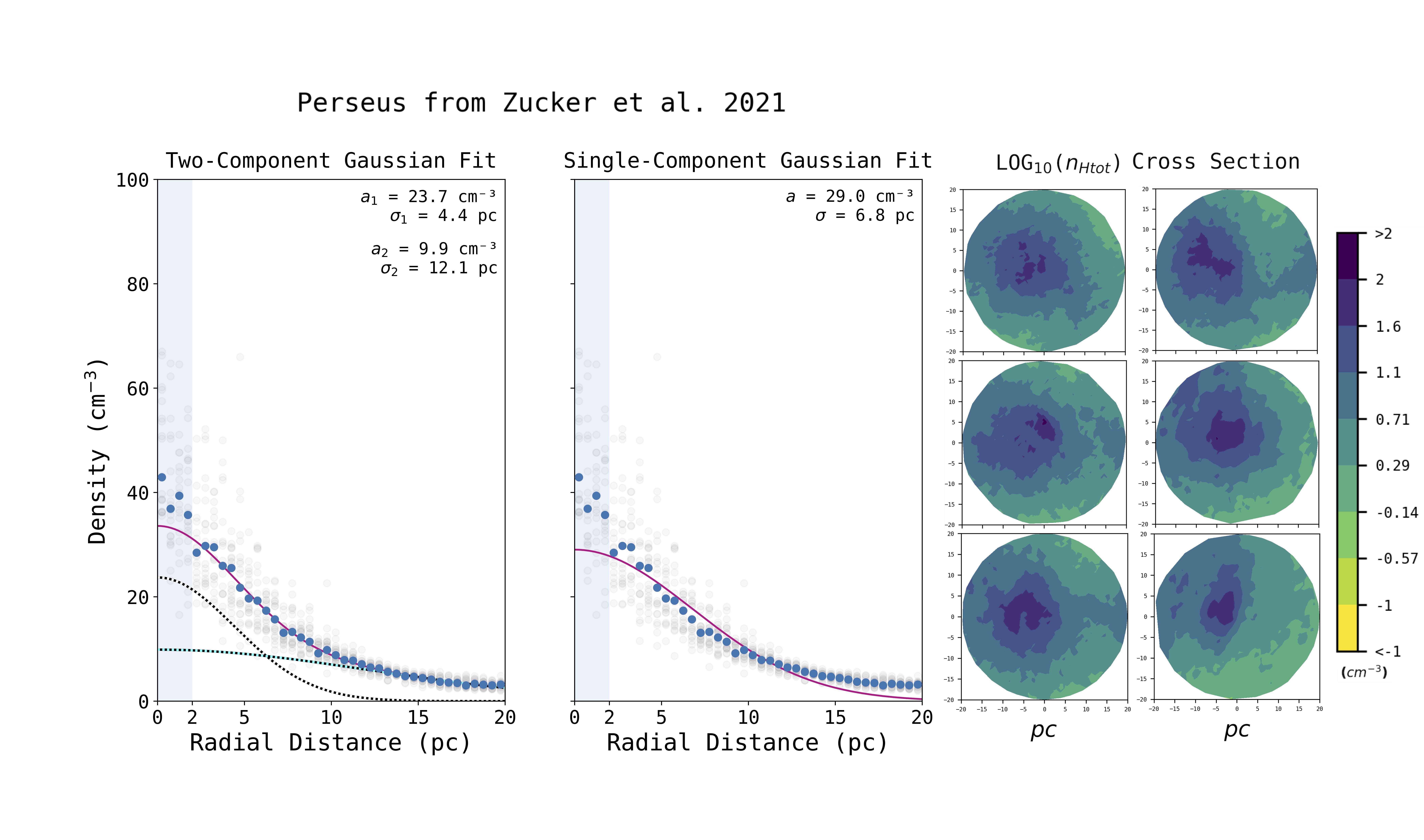}
\caption{Cloud averaged radial volume density ($n_{\rm H_{tot}}$) Gaussian fits for Perseus for the two-component (left panel) and single-component (middle panel) fits, shown alongside the density slices taken at six different positions along the skeleton (right panel). In the left and middle panels, the blue dots display the averaged radial profile, while the gray points indicate the variation in the radial profiles from slice to slice. The red line shows the best-fit function, which for the two-component Gaussian consists of inner (black dotted line) and outer (cyan dotted line) components. The inner two parsecs, where the 3D dust observations are unreliable, have been grayed out.  The best-fit values for the two-component Gaussian fit (the inner and outer Gaussian widths $\sigma_1$ and $\sigma_2$ and their corresponding amplitudes $a_1$ and $a_2$) and the single-component Gaussian fit (the width $\sigma$ and amplitude $a$) are summarized at top. As can be seen in the right panel, the slices display radial symmetry as well as an extended gas envelope.}
\label{Perseus Radial Profile}
\end{figure*}

\begin{figure*}[ht!]
\includegraphics[width=\linewidth]{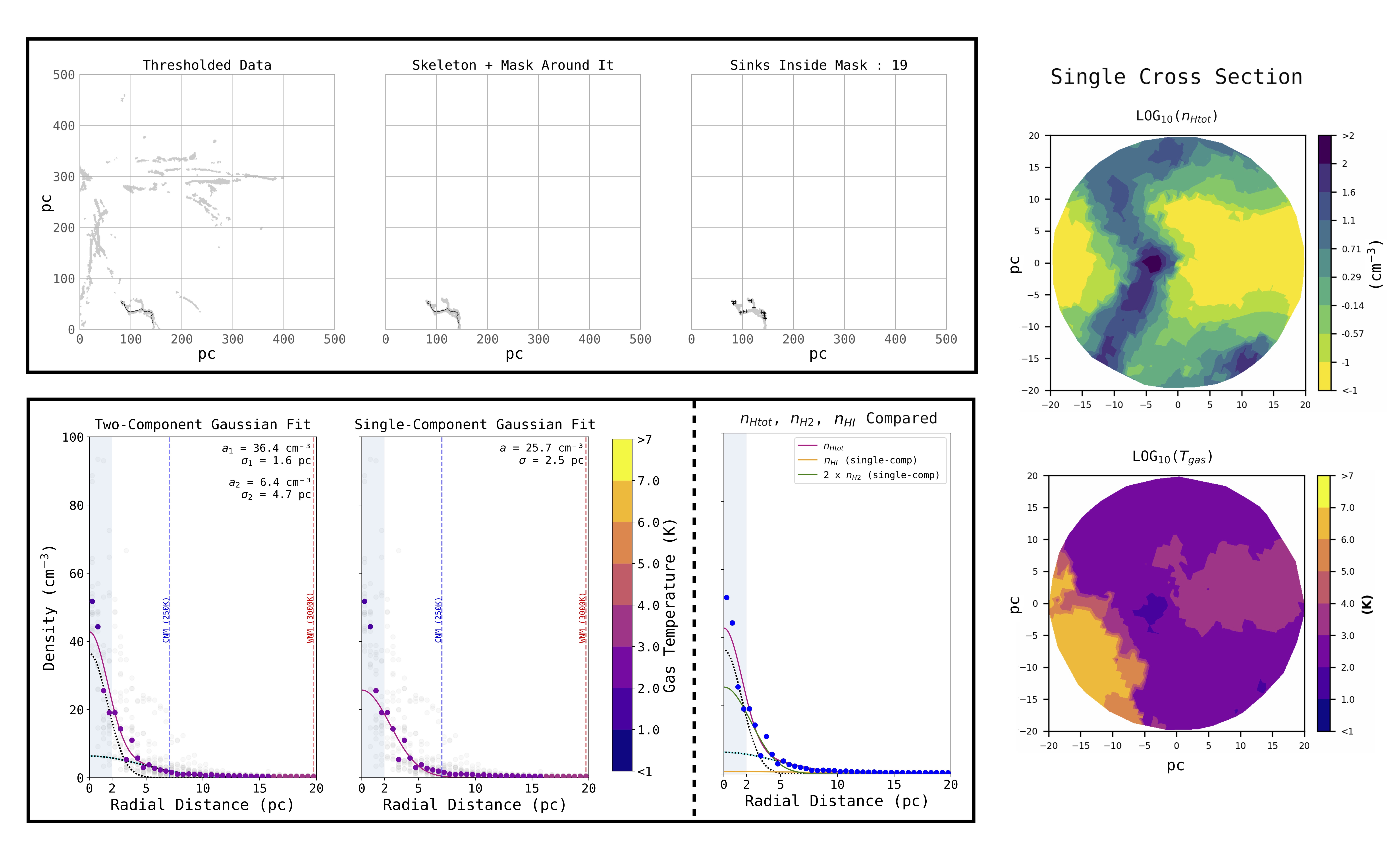}
\caption{Sheet-like simulated cloud (grid x1465y1400z1400, cloud 23) with nineteen star-forming sinks. In the top panel, we show the projected density field thresholded at a level of $n_{\rm H_{tot}}$ $\geq$ 35 cm$^{-3}$ (top left panel) alongside the cloud's mask and skeleton (top middle panel) and location of star-forming sinks (crossed points) within the cloud (top right panel). 
The bottom panel displays the cloud-averaged radial volume density ($n_{\rm H_{tot}}$) Gaussian fits for the simulated cloud with two-component (bottom, left panel) and single component (bottom, middle panel) fits. In the bottom left and bottom middle panels, the colored points display the averaged radial profile with the average gas temperature at that radial distance (colorbar), while the gray points indicate the variation in the radial profiles from slice to slice.  The red line shows the best-fit function, which for the two-component Gaussian consists of inner (black dotted line) and outer (cyan dotted line) components. The inner two parsecs, where the 3D dust observations (see Figure \ref{Perseus Radial Profile}) are unreliable, have been grayed out. The vertical dotted lines indicate temperature transitions from CNM to UNM ($\sim$250 K, blue dotted line) and UNM to WNM ($\sim$3000 K, red dotted line). The bottom right panel shows the $n_{\rm H_{tot}}$ two-component Gaussian fits (with the inner and outer components) compared to the single component Gaussian fits for $n_{\rm H_{2}}$ (orange) and $n_{\rm H_{I}}$ (green). The far right panels show representative density and temperature slices taken midway across the skeleton. As can be seen in the right panel, the slice morphology is much different than that found in Figure \ref{Perseus Radial Profile}.}
\label{Sheet}
\end{figure*}

\begin{figure*}[ht!]
\includegraphics[width=\linewidth]{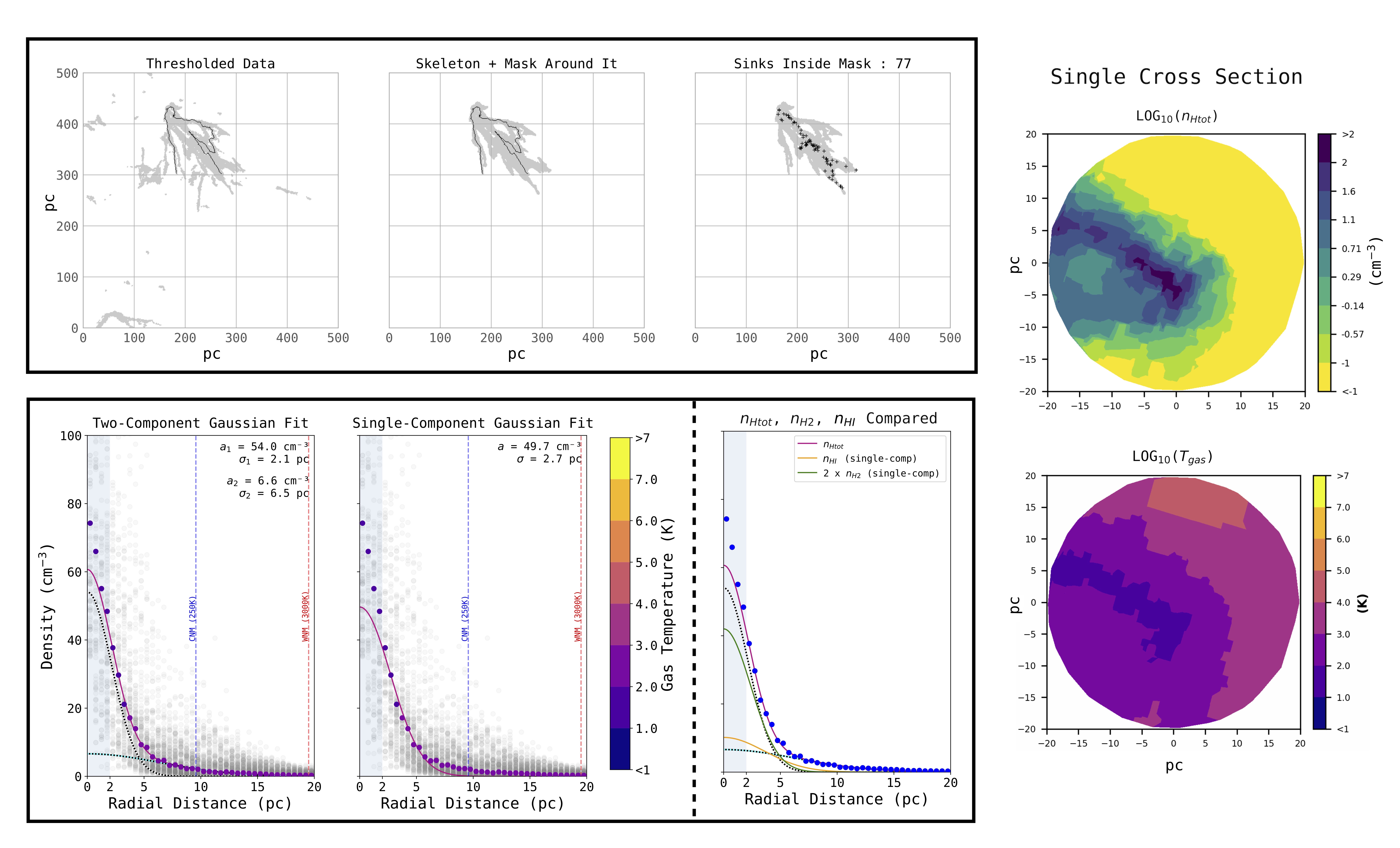}
\caption{Same as Figure \ref{Sheet}, but for an asymmetric cloud with 77 star-forming sinks. (grid x1465y1390z1400, cloud 0)}
\label{Big-Boy}
\end{figure*}

\begin{figure*}[ht!]
\includegraphics[width=\linewidth]{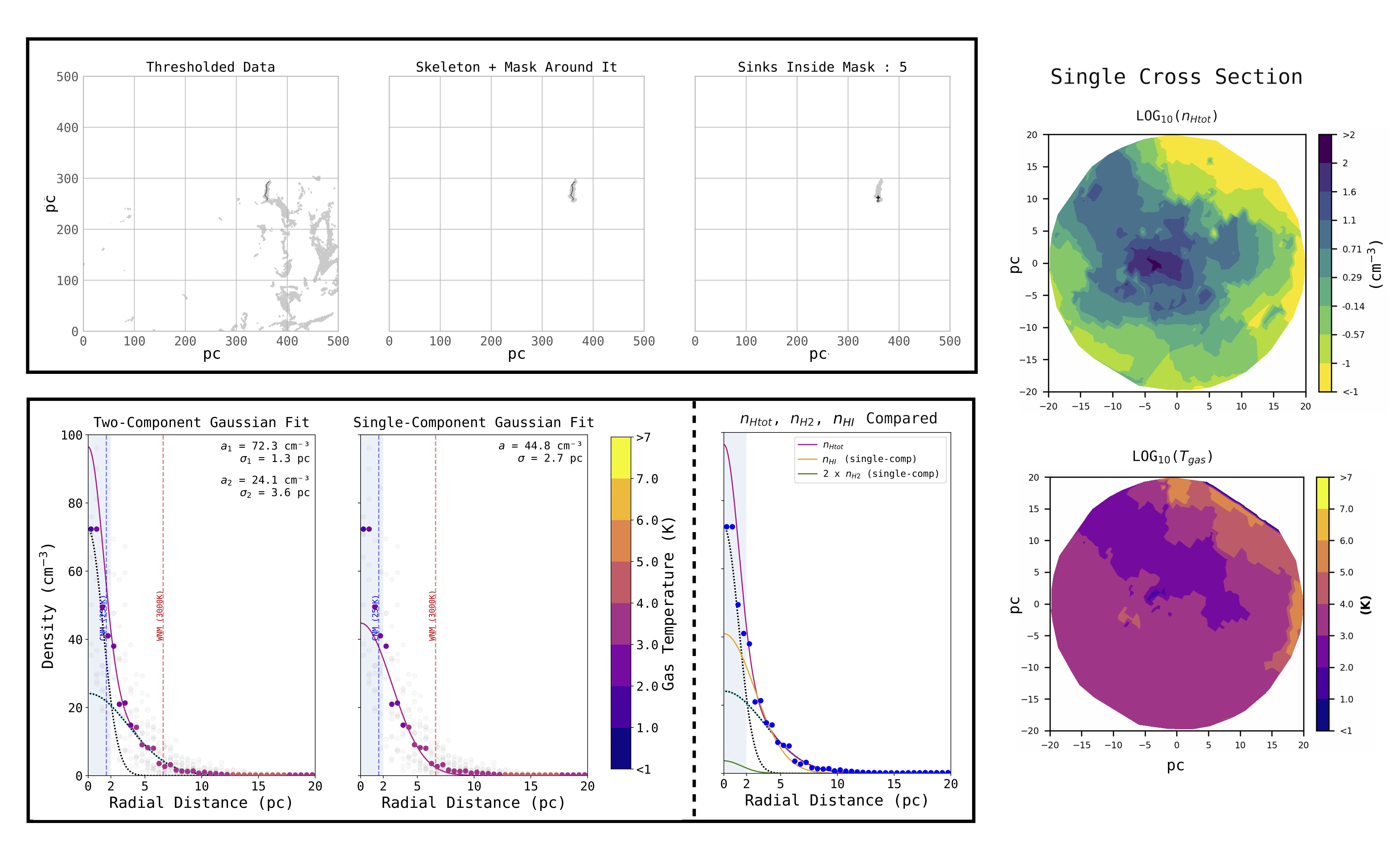}
\caption{Same as Figure \ref{Sheet}, but for a symmetric cloud with five star-forming sinks. (grid x14751390z1395, cloud 0)}
\label{Symmetric}
\end{figure*}

\begin{figure*}[ht!]
\centering
\includegraphics[width=\linewidth]{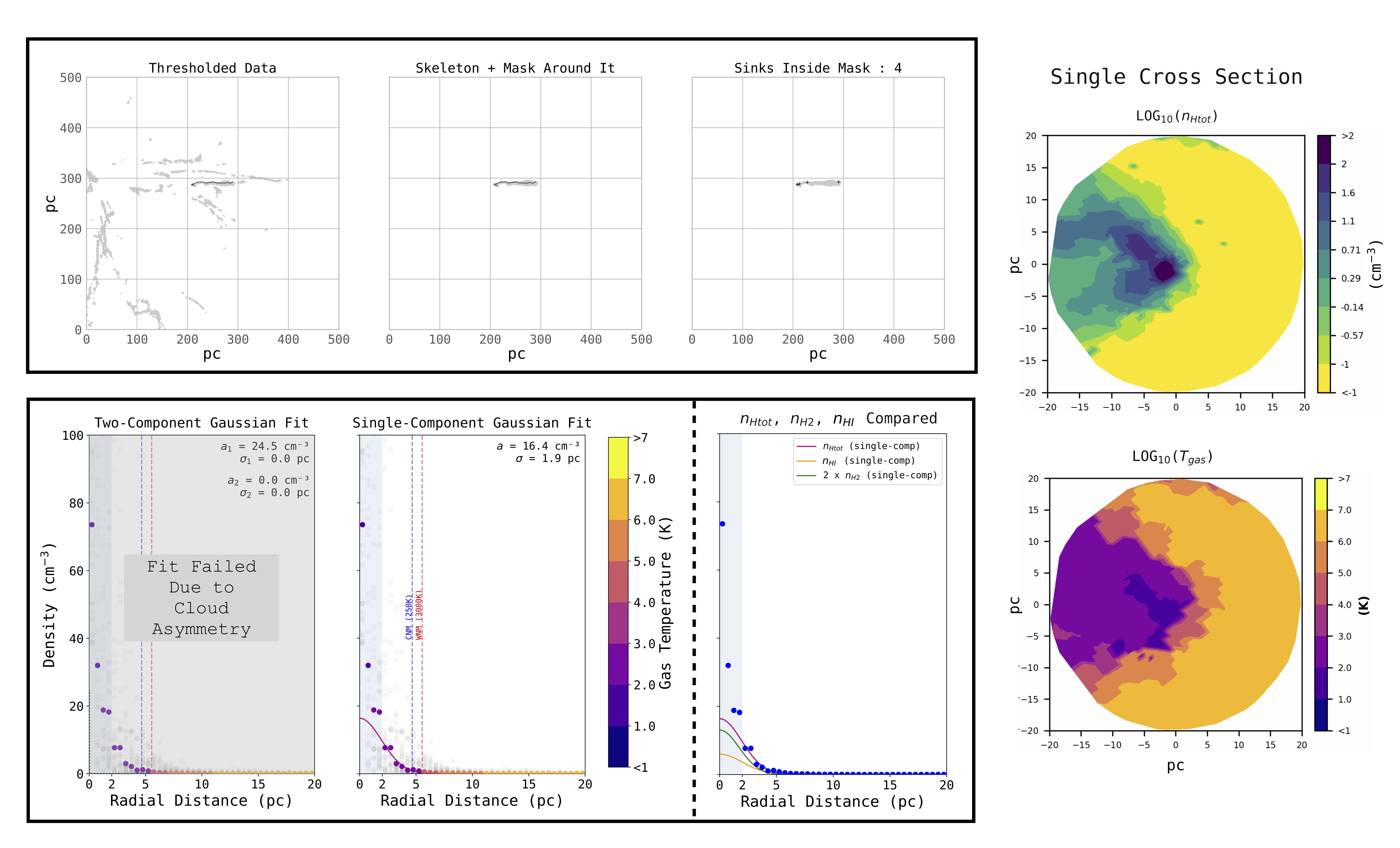}
\caption{Same as Figure \ref{Sheet}, but for an asymmetric cloud with four star-forming sinks. The two-component Gaussian fit failed due to an exponential-like radial profile. This failure likely stems from the assumption of radial symmetry in the derivation of the averaged radial profile. (grid x14651400z1400, cloud 8)}
\label{Failed}
\end{figure*}

\begin{figure*}[ht!]
\centering
\includegraphics[width=\linewidth]{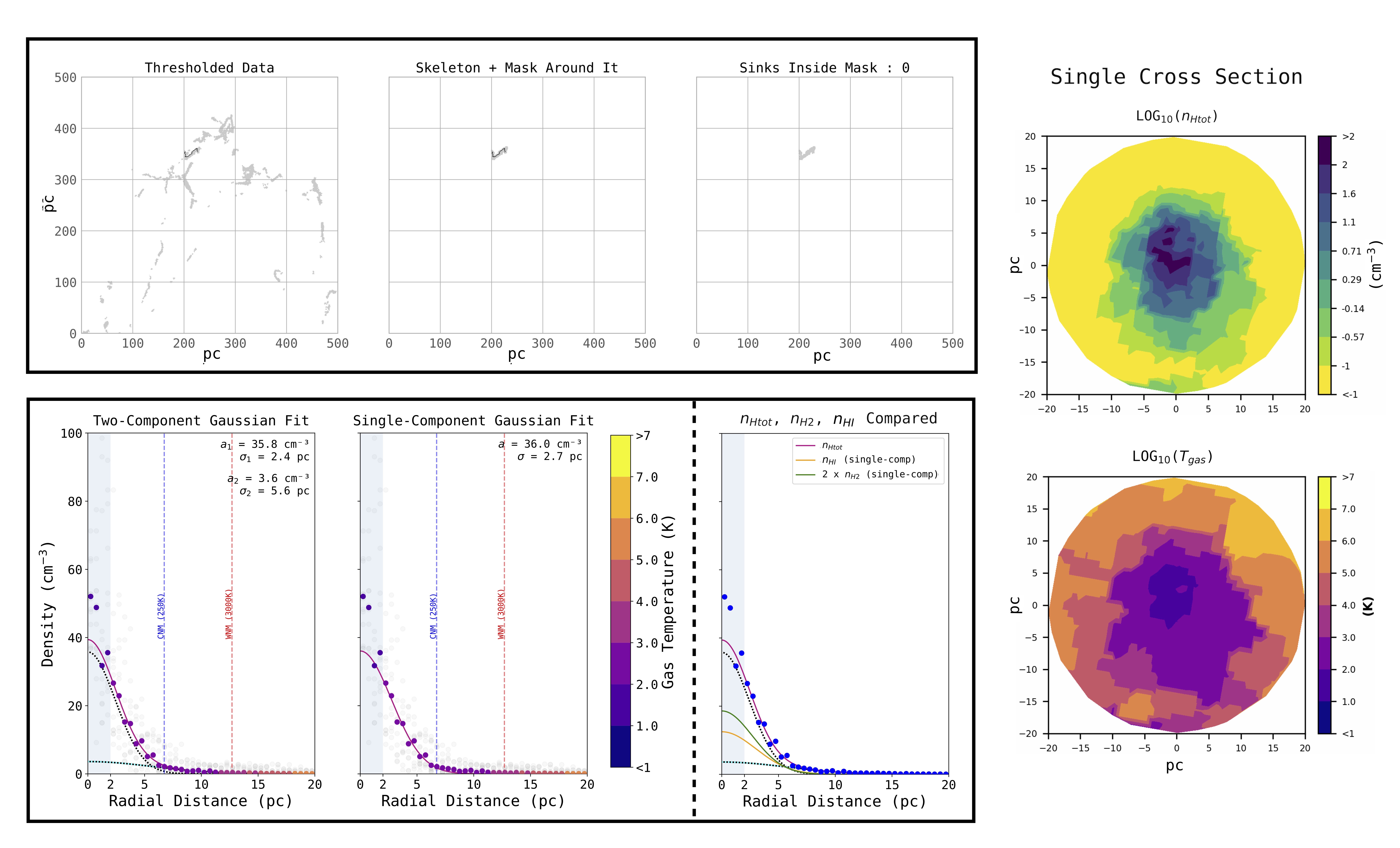}
\caption{Same as Figure \ref{Sheet}, but for a symmetric cloud with 0 star-forming sinks. Due to high density of molecular gas, this cloud could be in the precursor stages of star formation. (grid x14701400z1395, cloud 3)}
\label{No-Sinks}
\end{figure*}

As a basis for comparison with the observational data, we apply our pipeline to the local Perseus molecular cloud, whose radial density profile based on the 3D dust data from \citet{dust_maps} was originally analyzed in \citet{Zucker2021}. Figure \ref{Perseus Radial Profile} displays our pipeline's results for Perseus. The two radial profile plots display the two-component Gaussian and single-component Gaussian fit to Perseus’s radial profile. The gray points in the figure represent the slice-by-slice radial profiles that are averaged to compute the cloud-averaged radial volume density profile, represented by blue points since there is no temperature information present in the observational data. We also display multiple cross sections of the $n_{\rm H_{tot}}$ density of Perseus taken along the skeleton. Our pipeline reproduces a two-component Gaussian fit that traces the data better than the single-component fit. Discrepancies between our fitted values and the values found in \citet{Zucker2021} come from slightly different modeling choices and that our pipeline does not take into account errors intrinsic to observations (see \S \ref{sec:discussion}).

Figures \ref{Sheet}, \ref{Big-Boy}, and \ref{Symmetric} display three exemplary simulated star-forming clouds. The top panel displays the cloud's location in the simulation grid, the skeleton of the cloud, and the number and location of star-forming sinks within the cloud. The bottom panel displays the one- and two-component Gaussian fits to the $n_{\rm H_{tot}}$ radial profiles, where the temperature is displayed as a colorbar and dotted lines demarcate the radial distances corresponding to potential atomic gas thermal transition. Recall that the $n_{\rm H_{tot}}$ radial profile is the only information we can directly compare with observational data, as there is no observational information on the relative contribution from atomic versus molecular hydrogen gas. For the simulations we additionally include the radial profiles for molecular ($2 \times n_{\rm H_2}$) and atomic $n_{\rm H_I}$ gas densities alongside $n_{\rm H_{tot}}$. We also include two representative slices taken halfway through the cloud's skeletal length. These slices display total gas density $n_{\rm H_{tot}}$ and the gas temperature morphology. 

The three exemplary fits display a wide variety of morphologies. Figure \ref{Sheet} shows a sheet-like cloud, Figure \ref{Big-Boy} shows a radially asymmetric cloud, and Figure \ref{Symmetric} shows a cloud that is approximately radially symmetric. However these three fits display a common theme found in all clouds in the simulation: in general, the two-component Gaussian fit does not trace the data better than a single-component fit. Additionally, consistent with the small outer Gaussian width ($\sigma_2$) common across the entire ensemble of simulated clouds found in Figure \ref{fig:corner_Zucker}, the radial profiles for the example simulated clouds fall off to $\sim 0 \; \rm cm^{-3}$ more quickly than Perseus' profile (Figure \ref{Perseus Radial Profile}). This trend is even more evident in the representative slices, which all display a lack of an extended gas envelope. We find that the lack of an extended gas envelope is a common feature across the entire sample of clouds.

While simulated non-star-forming clouds cannot be directly used to disentangle the precise thermal and chemical phase structure of observed star-forming clouds analyzed in \citet{Zucker2021}, they can elucidate properties that are common within the entire simulation grid. The ensemble results including the entire sample size is shown in Appendix Figure \ref{fig:corner_zucker_full} and Table \ref{table:population_statistics}. The majority of non-star-forming clouds displays similar properties to the star-forming ones for all Gaussian fit parameters. In particular, the Gaussian widths ($\sigma_1,\sigma_2$), extend out to $\sim$2 pc, $\sim$5 pc for a majority of the simulated clouds. A small subset of non-star-forming clouds do display larger $\sigma_2$ widths ($\geq 10$ pc). These clouds, while not star forming, tend to be found in very dense regions of the simulation grid and provide an interesting subset of clouds that we defer to future investigation.

\section{Discussion}\label{sec:discussion}

From the slice visuals in the three exemplary fits, we determine that the poor two-component Gaussian fit to the $n_{\rm H_{tot}}$ radial profiles is due to radial asymmetry and lack of an extended envelope. All three figures (Figure \ref{Sheet}, \ref{Big-Boy}, \ref{Symmetric}) display a lack of an extended envelope and Figure \ref{Sheet}, \ref{Big-Boy} display radial asymmetry. The lack of extended envelope presents itself as a rapid falloff to $\sim$$0 \; \rm cm^{-3}$ at $\sim$10 pc in the cloud average radial volume density profiles. Radial asymmetry presents itself as a larger scatter of gray points in Figure \ref{Sheet} and \ref{Big-Boy}, with the gray points being representative of the individual slice-by-slice radial profiles that are averaged to compute the cloud averaged radial volume density profile. 

We turn to a more extreme example of asymmetry in a simulated star-forming cloud in Figure \ref{Failed}. This cloud displays radial asymmetry and lacks an envelope to such an extreme extent that the two-component Gaussian fit fails to converge. Asymmetry in this case causes the radial profile to take on a more exponential shape than a Gaussian profile. We can compare the density slice of this cloud to the density slices for Perseus. Perseus' density slices display approximate radial symmetry and a gas envelope that extends past 20 pc. The Log($T_{\rm gas}$) slices map onto the Log($n_{\rm H_{tot}}$) slices quite well and we identify a temperature front that can exceed $\sim 10^6 \; \rm K$ mapping onto regions of the cloud with $\sim 0.001 \; \rm cm^{-3}$ density. In the Cloud Factory simulations there is a constant UV background that acts as a source of background heating. The resultant temperature field is driven by density variations where lower density gas is hotter due to being less shielded from the interstellar radiation field. This example is in agreement with the asymmetric cloud in Figure \ref{Big-Boy}. This cloud also displays a temperature front that maps onto the density structure, albeit a much colder one due to the skeleton tracing a part of the cloud that is being shielded. Similar temperature fronts map to regions of lower density in both Figure \ref{Sheet} and Figure \ref{Symmetric}.

The lack of an extended gas envelope could be due to two factors: 1) too strong supernova feedback in comparison to the solar neighborhood or 2) a lack of magnetic fields which would help keep clouds together \citep[see e.g., discussion in][]{Shashwata2023}. Supernova feedback injects both thermal energy and momentum into the ISM and could be the cause of the low density envelopes, and therefore the low $\sigma_2$ values. To probe for a potential cause we repeated the experiment on one grid of the ``potential-dominated" suite of the Cloud Factory simulations, where supernova feedback is less prominent and the large-scale gravitational potential largely dictates the gas dynamics. The resultant cloud profiles were predominately sheet-like and displayed the same lack of extended gas envelopes.  We display 2D representations of a feedback-dominated and potential-dominated grid, as well as 6 different cloud morphologies from both grids, in Appendix Figure \ref{fig:feedback_vs_potential}. This result gives credence to magnetic fields being responsible for the lack of extended envelopes; however we leave a more detailed comparison between the feedback and potential dominated grids to future studies. 

Our finding is in agreement with the work performed by \citet{Shashwata2023}. \citet{Shashwata2023} utilized the SILCC-Zoom simulations \citep{Seifried2017}, which include magnetic fields, self-gravity, supernova feedback, and non-equilibrium chemistry to study the effect of magnetic fields on molecular clouds. They found that molecular clouds in their subset of simulations with magnetic fields contained more mass in their diffuse envelopes than clouds without. They additionally found that magnetic fields as a whole are more important for less dense structures. This result provides evidence that the lack of magnetic fields in the Cloud Factory suite may be driving the lack of extended envelopes in simulated clouds.

To summarize: clouds in the ``feedback dominated" suite of the Cloud Factory simulations display a lack of extended envelope, a finding also shared by a subset of clouds in the ``potential dominated" suite. Star-forming clouds in the ``feedback dominated" case can be radially asymmetric filaments (Figure \ref{Big-Boy}, \ref{Failed}), symmetric filaments (Figure \ref{Symmetric}), or sheet-like (Figure \ref{Sheet}) with a majority of the clouds being radially asymmetric. Star-forming clouds in the ``potential dominated" case are solely sheet-like. We propose that the lack of magnetic fields in combination with the constant UV background radiation is driving the lack of low-density envelopes in both simulation suites. That is, the lack of envelopes in both suites is due to clouds not being able to hold themselves together against the constant UV radiation field, thereby destroying the extended low-density envelopes. Additionally we argue that a combination of the galactic potential and supernova feedback are driving the cloud morphologies; the radially asymmetric profiles in the ``feedback dominated" suite is due to supernova feedback dominating, while sheet-like structures is due to potential forces dominating.

Our results are independent of modeling choices. We revisited the choice of restricting our analysis to clouds with lengths greater than 20 pc (\S \ref{sec:Methods}) by rerunning the pipeline with the opposite criterion (restricting the analysis to clouds less than 20 pc) for a single feedback-dominated grid. We find that these small simulated clouds have similar morphologies to the longer clouds but --- due to their radial profiles only having twenty slices --- the shorter clouds displayed significantly more scatter in their averaged profile. We additionally analyzed the effect of including a `shift' of the cloud center from the topological center to the densest region within the cloud's mask, ensuring that the density peaks at a distance of 0 pc. This `shift' was explored in both \citet{Zucker2018} and \citet{Zucker2021}. Including this shift does not impact our main result, as the effect of slice-by-slice differences and cloud morphology impacts the resultant radial profile more profoundly. We note that shifting remains an option in the pipeline, and can be pertinent when specifically analyzing peak densities and width values in radial profiles of observations and simulations.

We finally revisit our assumption of radial symmetry when extracting radial profiles from each 20 pc slice. Due to the diversity of cloud morphologies found in the simulations, assuming radial symmetry is not appropriate for asymmetric and sheet-like clouds and new methods must be utilized to examine the density structure of these clouds. Departing from radial symmetry is outside of the scope of this work, and is why the pipeline provides slices as a tool to analyze whether radial symmetry is a good assumption or not. 

In comparison to this work, \citet{Zucker2021} performed Bayesian model comparison to show that a two-component Gaussian fit was preferred over other models, including a single-component Gaussian fit. Given that the simulations do not match observations, similar Bayesian model comparison for the simulated cloud radial profiles is beyond the scope of this work. However, we note that the pipeline we present here should have broad potential applications, and more sophisticated Bayesian modeling of the extracted radial profiles can be applied to future simulations that include additional physics. 

As shown in Figure \ref{Sheet}, \ref{Big-Boy}, \ref{Symmetric}, \ref{Failed}, \ and Appendix Figure \ref{fig:forward_model}, the total gas density, chemical composition, and temperature radial profile slices map onto each other quite well. In the future, if simulated clouds better reproduce the extended envelopes seen in nearby observed clouds, radial profile slices of both temperature and other density tracers can be better inferred from the total gas density profiles. We would then be able to extrapolate additional information of observed clouds from their total gas density, and test the radial distances at which local molecular filamentary clouds may exhibit thermal and/or chemical phase transitions. 

We turn to a final example in Fig \ref{No-Sinks}. This cloud has a symmetric radial morphology and contains no star-forming sinks. This cloud is still noteworthy within the broader sample since even though it is not star-forming, it contains more molecular gas in the inner few parsecs than atomic gas. This cloud could be in the precursor stages of star-formation. Tracking the evolution of morphology and chemical composition of clouds like these over time can elucidate the life cycle of a molecular cloud as it transitions from non-star-forming to star-forming, which we defer to future work. 

\section{Conclusion} \label{sec:conclusion}

We present a technique to extract and analyze clouds in resolved simulations of the ISM for the purpose of comparing to 3D observational data. We first extend the cloud-identification and skeltonization methodology presented in \citet{Zucker2021} to search simulation grids for cloud-like structures. The radial profiles of clouds along their topological skeletons are then computed in multiple tracers. Density radial profiles are fit using one-component and two-component Gaussians and temperature radial profiles are fit using sigmoid and spline profiles. 2D slices along the skeleton are also computed to serve as a visualization of cloud morphology.

We apply the pipeline to the Cloud Factory feedback-dominated suite of simulations to extract radial profiles of two dozen star-forming clouds. We find that simulated clouds are primarily radially asymmetric and lack low-density envelopes. In line with recent work \citep[e.g.,][]{Shashwata2023}, we propose that the lack of low-density envelopes is due to the lack of magnetic fields. We additionally propose that radially asymmetry morphologies is due to supernova feedback. Initial tests indicate that clouds in the Cloud Factory potential-dominated suite of simulations are primarily sheet-like and also lack low-density envelopes, giving credence to the common lack of magnetic fields in both suites driving the lack of a low-density envelope and supernova feedback driving radially asymmetric morphologies. Future work will investigate the differences between the feedback and potential-dominated suites in more detail, and compare the Cloud Factory clouds to clouds in other simulations. 

New observational techniques that probe the 3D structure of local star-forming clouds provides an exciting new avenue by which simulations can be used to detangle the physical conditions occurring in observed clouds. We show that there is a current discrepancy between the 3D structure of clouds in hydrodynamic simulation and observations. Future iterations of the Cloud Factory suite will include magnetic fields \citep[as in e.g.,][]{Shashwata2023} and track cloud formation and evolution over time. On the observational front, there are new 3D dust maps being developed \citep[e.g.][]{Gordian2023} that extend to larger distances. These new 3D dust maps will  allow for opportunities to characterize cloud morphologies over a more diverse range of galactic environments. The cloud identification and analysis pipeline we present here provides a tool for future exploration of simulated and observed 3D cloud structure and is publicly available via \href{https://doi.org/10.5281/zenodo.10157333}{Zenodo (doi: 10157333)} and \href{https://github.com/elijah-mullens/Skeletonizing-and-Analyzing-Pipeline-for-3D-Interstellar-Cloud-Ensembles}{GitHub (Skeletonizing and Analyzing Pipeline for 3D Interstellar Cloud Ensembles)}.

\hfill

We acknowledge support from Space Telescope Science Institute's 2022 Space Astronomy Summer Program. We thank STScI's continued support of this project by funding two in-person poster presentations at the 241st meeting of the American Astronomical Society and Protostars and Planets VII. E.M. acknowledges that this material is based upon work supported by the National Science Foundation Graduate Research Fellowship under Grant No.\ 2139899. We thank the anonymous referee for a very constructive and thorough report of our manuscript that significantly improved the quality of this work. 

\clearpage
\appendix

\begin{turnpage}

\begin{deluxetable}{lRRRRRRRRRRRRRRRR}
\tabletypesize{\footnotesize}
\tablecolumns{17}
\tablewidth{0pt}
\colnumbers
\tablecaption{Properties of Simulated Star-Forming Clouds \label{table:everything}}
\tablehead{
Grid \# &  \rm{Cloud \; \#} & \rm{Length} &  \rm{Mass} & \rm{\# \; Sinks} & \multicolumn{6}{|c|}{$n_{\rm H_{tot}}$} &  \multicolumn{2}{|c|}{$n_{\rm CO}$} &  \multicolumn{2}{|c|}{$n_{\rm H_2}$} & \multicolumn{2}{|c|}{$n_{\rm H_I}$} \\
 &  &  &  &  &  $a_1$ &  $\sigma_1$&  $a_2$&  $\sigma_2$&  $a$ &  $\sigma$  & $a$ &  $\sigma$ &  $a$ & $\sigma$&  $a$ &  $\sigma$\\
 & &  \rm{pc} &  \rm{M$_\odot$} &  &  \rm{cm^{-3}} & \rm{pc} & \rm{cm^{-3}} & \rm{pc} & \rm{cm^{-3}} & \rm{pc} & \rm{cm^{-3}} & \rm{pc} & \rm{cm^{-3}} & \rm{pc} & \rm{cm^{-3}} & \rm{pc} \\}
 \startdata
x1465y1390z1400 &        0 &          545 &        322558 &       77 &              54.0 &                 2.1 &               6.6 &                 6.5 &             50.0 &               2.7  & 1.9\text{e-}06 &             1.0 &           21.0 &             2.4 &           10.0 &             1.0 \\
x1465y1390z1400 &        3 &           87 &         18735 &        9 &              54.0 &                 1.0 &              13.0 &                 3.6 &             16.0 &               3.3 & 2.9\text{e-}09 &             1.7 &            4.6 &             3.2 &            7.2 &             1.7  \\
x1465y1390z1400 &        4 &          225 &         37483 &       15 &              69.0 &                 1.1 &              18.0 &                 3.9 &             27.0 &               3.2 & 1.3\text{e-}07 &             1.5 &           12.0 &             3.0 &            2.4 &             1.5  \\
x1465y1390z1400 &       12 &          191 &         25118 &        1 &              42.0 &                 1.9 &               3.2 &                 5.9 &             38.0 &               2.2 & 6.1\text{e-}09 &             1.2 &            3.8 &             1.8 &           31.0 &             1.2 \\
x1470y1400z1395 &        4 &           76 &          5812 &        1 &              23.0 &                 2.6 &               9.5 &                 2.6 &             33.0 &               2.6 & 3.6\text{e-}09 &             1.6 &           10.0 &             2.2 &           19.0 &             1.6  \\
x1465y1405z1395 &        8 &          169 &         63898 &        3 &              58.0 &                 2.0 &               3.1 &                 7.1 &             54.0 &               2.2 & 2.4\text{e-}08 &             1.2 &            3.8 &             1.7 &           47.0 &             1.2 \\
x1475y1390z1395 &        0 &           46 &          6882 &        5 &              72.0 &                 1.3 &              24.0 &                 3.6 &             45.0 &               2.7 & 5.7\text{e-}06 &             0.6 &            1.8 &             1.5 &           41.0 &             0.6 \\
x1475y1390z1395 &        2 &          101 &         19255 &        2 &              79.0 &                 1.5 &              19.0 &                 4.9 &             44.0 &               3.1 & 6.6\text{e-}07 &             1.0 &           12.0 &             2.3 &           24.0 &             1.0 \\
x1475y1390z1395 &        4 &          313 &         75488 &        3 &              32.0 &                 2.5 &              13.0 &                 5.9 &             36.0 &               3.7 & 4.4\text{e-}07 &             1.1 &            7.2 &             3.3 &           17.0 &             1.1\\
x1475y1390z1395 &        8 &           37 &          5176 &        1 &              30.0 &                 2.0 &               1.7 &                 7.1 &             28.0 &               2.3 & 5.1\text{e-}09 &             1.7 &           12.0 &             2.0 &            7.0 &             1.7  \\
x1475y1390z1395 &       18 &           18 &         48825 &       19 &              57.0 &                 1.7 &               9.0 &                 4.7 &             44.0 &               2.4 & 9.8\text{e-}04 &             0.6 &           21.0 &             2.2 &            3.3 &             0.6 \\
x1475y1390z1395 &       22 &          122 &         19959 &        8 &              54.0 &                 1.4 &              12.0 &                 3.8 &             49.0 &               1.9 & 2.2\text{e-}04 &             0.5 &          540.0 &             0.7 &           19.0 &             0.5\\
x1465y1400z1400 &        3 &           33 &         15656 &        5 &              86.0 &                 2.1 &               9.8 &                 5.9 &             80.0 &               2.6 & 5.8\text{e-}03 &             1.7 &            0.0 &             2.2 &           80.0 &             1.7\\
x1465y1400z1400 &       13 &           43 &          3039 &        1 &              48.0 &                 1.2 &              14.0 &                 2.9 &             29.0 &               2.2 & 6.5\text{e-}08 &             0.6 &            3.3 &             1.1 &           23.0 &             0.6 \\
x1465y1400z1400 &       18 &           69 &         10210 &        3 &              40.0 &                 1.9 &               3.3 &                 6.3 &             35.0 &               2.3 & 4.6\text{e-}09 &             1.4 &           13.0 &             1.8 &           16.0 &             1.4\\
x1465y1400z1400 &       23 &          125 &         23166 &       19 &              36.0 &                 1.6 &               6.4 &                 4.7 &             26.0 &               2.5 & 6.0\text{e-}08 &             1.1 &           13.0 &             2.4 &            0.7 &             1.1\\
x1470y1400z1400 &        1 &           36 &          2294 &        1 &              49.0 &                 1.7 &               6.1 &                 1.7 &             55.0 &               1.7 & 1.8\text{e-}07 &             0.8 &           34.0 &             1.3 &           10.0 &             0.8 \\
x1465y1405z1400 &        0 &          207 &         47893 &        8 &              78.0 &                 1.4 &               9.3 &                 4.4 &             47.0 &               2.2 & 3.0\text{e-}07 &             0.8 &            9.5 &             1.9 &           26.0 &             0.8\\
x1465y1405z1400 &        4 &          128 &         33625 &        2 &              69.0 &                 1.9 &               6.3 &                 8.7 &             56.0 &               2.5 & 2.1\text{e-}07 &             1.8 &           12.0 &             2.0 &           31.0 &             1.8  \\
x1465y1405z1400 &        8 &           21 &          1306 &        1 &              45.0 &                 1.7 &               0.9 &                 6.5 &             43.0 &               1.8 & 2.1\text{e-}08 &             1.0 &           17.0 &             1.8 &            4.5 &             1.0 \\
x1465y1405z1400 &       14 &           48 &         37131 &        7 &              27.0 &                 3.2 &               9.3 &                 3.2 &             36.0 &               3.2 & 1.1\text{e-}06 &             1.2 &            6.3 &             3.2 &           21.0 &             1.2 \\
x1465y1405z1400 &       18 &           34 &          4622 &        1 &              15.0 &                 1.2 &               6.1 &                 5.0 &              7.4 &               4.5 & 3.8\text{e-}10  &             2.0 &            1.1 &             3.8 &            4.6 &             2.0\\
x1465y1405z1400 &       28 &           41 &          4239 &        3 &              36.0 &                 1.7 &               5.7 &                 5.7 &             26.0 &               2.6 & 3.2\text{e-}09 &             1.4 &            8.1 &             2.3 &            8.9 &             1.4 \\
x1475y1390z1400 &        5 &           96 &          5630 &        1 &              30.0 &                 2.1 &               5.3 &                 4.2 &             30.0 &               2.5 & 9.5\text{e-}09 &             1.1 &            6.2 &             2.4 &           19.0 &             1.1 \\
\enddata
\tablecomments{Skeletonizing and analysis pipeline results for star-forming clouds in the Cloud Factory feedback-dominated suite of simulations. (1) Grid number identifier. (2) Cloud number identifier (each skeletonized cloud in one simulation grid is given an identifying number). (3) Length and (4) Mass of each cloud. (5) Number of star-forming sinks in each cloud. Two-component Gaussian fit (amplitudes $a_1,a_2$ and standard deviation $\sigma_1,\sigma_2$) results to the (6-9) $n_{\rm H_{tot}}$ radial profiles. One-component Gaussian fit (amplitude $a$ and standard deviation $\sigma$) results to the (10-11) $n_{\rm H_{tot}}$, (12-13) $n_{\rm CO}$, (14-15) $n_{\rm H_{2}}$, and (16-17) $n_{\rm H_{I}}$ radial profiles.}
\end{deluxetable}

\begin{deluxetable}{lRRRRRRRRRRRRRR}
\tabletypesize{\footnotesize}
\colnumbers
\tablecolumns{15}
\tablewidth{0pt}
\tablecaption{Population Statistics \label{table:population_statistics}}
\tablehead{
 Type & \rm{Length} &  \rm{Mass}  & \multicolumn{6}{|c|}{$n_{\rm H_{tot}}$} &  \multicolumn{2}{|c|}{$n_{\rm CO}$} &  \multicolumn{2}{|c|}{$n_{\rm H_2}$} & \multicolumn{2}{|c|}{$n_{\rm H_I}$} \\
 &  &  &  $a_1$ &  $\sigma_1$&  $a_2$&  $\sigma_2$&  $a$ &  $\sigma$  & $a$ &  $\sigma$ &  $a$ & $\sigma$&  $a$ &  $\sigma$\\
 & \rm{pc} &  \rm{M$_\odot$} & \rm{cm$^{-3}$} & \rm{pc} & \rm{cm$^{-3}$} & \rm{pc} & \rm{cm$^{-3}$} & \rm{pc} & \rm{cm$^{-3}$} & \rm{pc} & \rm{cm$^{-3}$} & \rm{pc} & \rm{cm$^{-3}$} & \rm{pc} \\}
\startdata
Simulated Star-Forming & 81.5_{-46.1}^{+114.6} & 1.9e4_{-1.4e4}^{+2.9e4} & 48.5_{-18.5}^{+21.5} & 1.7_{-0.4}^{+0.4} & 7.8_{-4.5}^{+5.5} & 4.8_{-1.3}^{+1.7} & 37.0_{-10.3}^{+14.3} & 2.5_{-0.3}^{+0.7} & 9.8\text{e-}8_{-9.3\text{e-}8}^{+3\text{e-}6}& 1.2_{-0.4}^{+0.6} & 9.8_{-6.1}^{+8.5} & 2.2_{-0.6}^{+0.9} & 18.0_{-13.4}^{+13.0} & 1.2_{-0.4}^{+0.6}\\
Simulated Non-Star Forming  & 36.0_{-16.0}^{+45.0} & 2.3e3_{-1.3e3}^{+6.2e3} & 38.0_{-13.0}^{+18.0} & 1.8_{-0.5}^{+0.6} & 9.2_{-5.6}^{+6.6} & 4.3_{-1.6}^{+2.8} & 34.0_{-11.8}^{+14.0} & 2.7_{-0.6}^{+0.5} & 5.5\text{e-}9_{-5.1\text{e-}9}^{+8.1\text{e-}8} & 1.1_{-0.4}^{+0.6} & 3.3_{-2.5}^{+7.5} & 1.9_{-0.5}^{+0.6} & 27.0_{-12.7}^{+9.5} & 1.1_{-0.4}^{+0.6}\\
Observations & 62.0_{-20}^{+14} & 9.4e3_{-6.4e3}^{+6.4e3} & 31.3_{-2.0}^{+10.9} & 2.9_{-0.3}^{+1.2} & 9.2_{-0.4}^{+2.4} & 10.9_{-1.7}^{+2.6} & 30.8_{-1.7}^{+10.2} & 5.7_{-1.3}^{+0.4} & --  & --  & --  & --  & --  & -- \\
\enddata
\tablecomments{Summary table of the ensemble median cloud properties and their dispersions for the simulated star-forming and non star-forming populations compared with observations. Each cell displays the median value (50th percentile) with the lower (50th percentile $-$ 16th percentile) and upper (84th percentile $-$ 50th percentile) bounds. (1) Simulation type. (2) Length and (3) Mass of the clouds. Two-component Gaussian fit (amplitudes $a_1,a_2$ and standard deviation $\sigma_1,\sigma_2$) results to the (4-7) $n_{\rm H_{tot}}$ radial profiles. One-component Gaussian fit (amplitude $a$ and standard deviation $\sigma$) results to the (8-9) $n_{\rm H_{tot}}$, (10-11) $n_{\rm CO}$, (12-13) $n_{\rm H_{2}}$, and (14-15) $n_{\rm H_{I}}$ radial profiles}
\end{deluxetable}

\end{turnpage}


\begin{figure}[h!]
\centering
\includegraphics[width=1\textwidth]{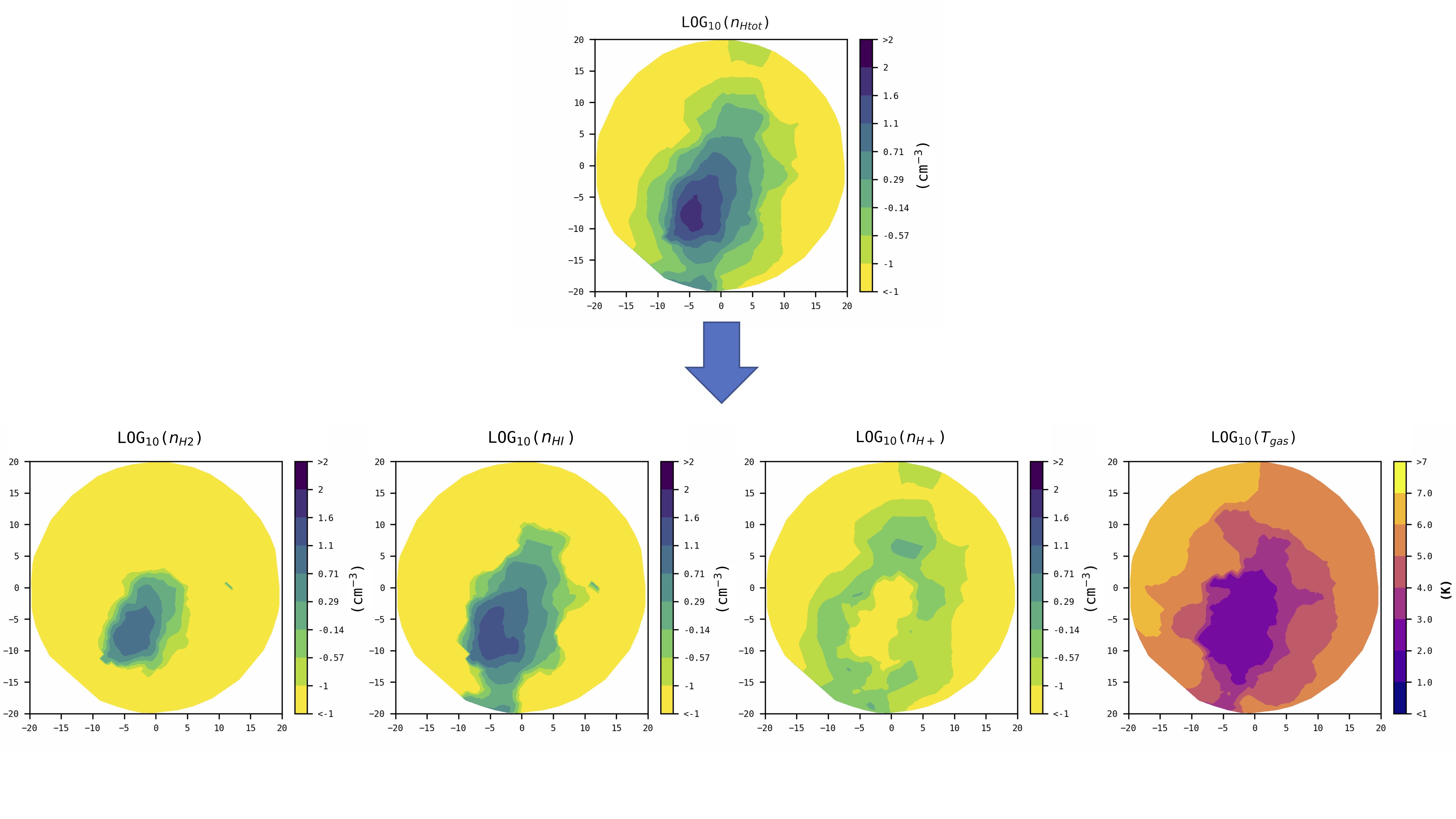}
\caption{Example slices taken from multiple tracers ($n_{\rm H_{tot}}$, $n_{\rm H_2}$, $n_{\rm H_I}$, $n_{\rm H_+}$, and $T_{\rm gas}$) midway along a skeleton. Once simulations can more faithfully reproduce the extended structure of gaseous envelopes observed in the solar neighborhood, we should be able to map these tracers on to the total hydrogen volume density to determine the radial distances at which chemical and/or phase transitions occur.}
\label{fig:forward_model}
\end{figure}

\begin{figure}[h!]
\centering
\includegraphics[width=1\textwidth]{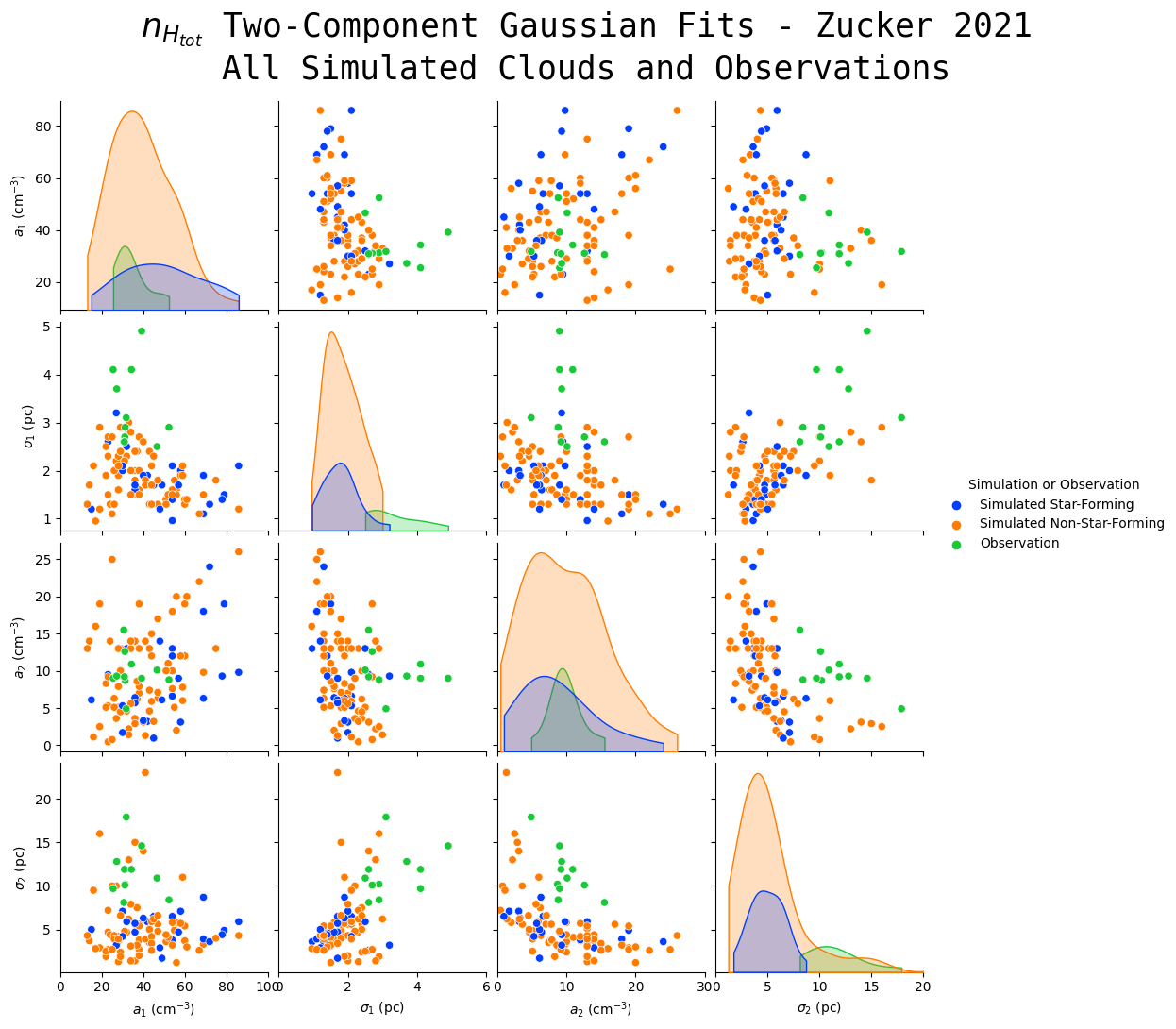}
\caption{Same as in Figure \ref{fig:corner_Zucker}, but with the inclusion of clouds with no star-forming sinks.}
\label{fig:corner_zucker_full}
\end{figure}

\begin{figure}[h!]
\centering
\includegraphics[width=1\textwidth]{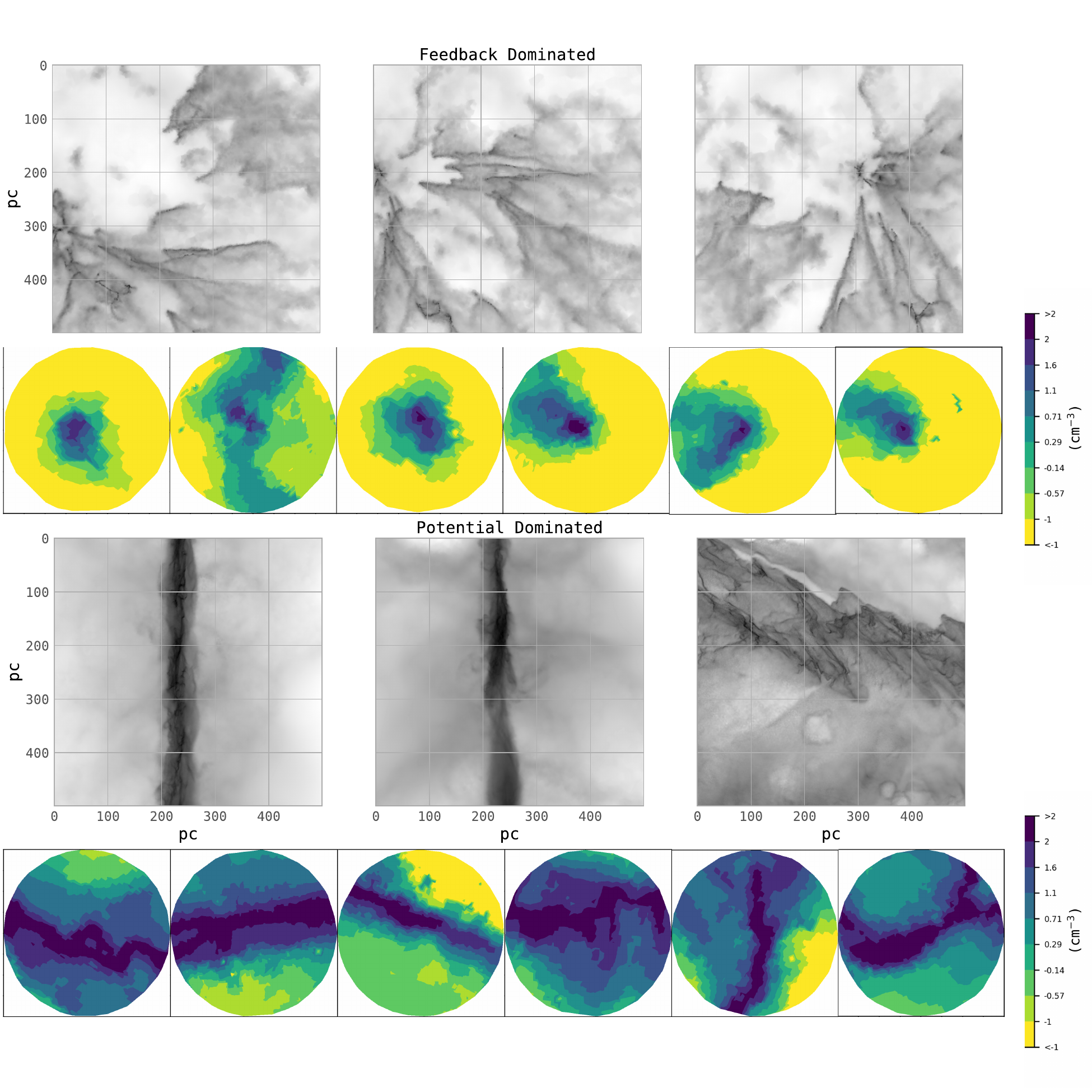}
\caption{Figure displaying the difference between the feedback (grid x1465y1400z1400) and potential (grid x1480y1390zp) dominated grids, both 500 pc $\times$ 500 pc in size. Panels display projected two-dimensional representations of the three-dimensional density grids by taking a sum along one axis. The potential-dominated grid contains a disk of high-density gas compared to the feedback-dominated grid. Due to the higher density, in order to run the pipeline, the mask was applied at a thresholded density of 500 cm$^-3$ in lieu of the 35 cm$^-3$ used in the feedback dominated case. The slices (20 pc $\times$ 20 pc in size) display the morphology of clouds found in each grid. The clouds in the potential dominated case are more sheet-like and denser than clouds in the feedback dominated case.}
\label{fig:feedback_vs_potential}
\end{figure}

\clearpage
\bibliography{sample631}{}
\bibliographystyle{aasjournal}

\end{document}